 \documentclass[manuscript]{aastex}

\shorttitle{Cosmic Diffuse Ultraviolet Background Radiation}
\shortauthors{Henry et al.}
\begin{document}
\title{The Mystery of the \\Cosmic Diffuse Ultraviolet Background Radiation}
\author{Richard Conn Henry }
\affil{Henry A. Rowland Department of Physics and Astronomy, \\The Johns Hopkins University,
    Baltimore, MD 21218, USA}
\email{henry@jhu.edu}

\author{Jayant Murthy}
\affil{Indian Institute of Astrophysics, Bengaluru, India}
\email{jmurthy@yahoo.com}

\author{James Overduin}
\affil{Department of Physics, Astronomy \& Geosciences, Towson University, Towson, MD 21252, USA}
\email{joverduin@towson.edu}

\author{Joshua Tyler\altaffilmark{1}
\affil{Department of Physics, Astronomy \& Geosciences, Towson University, Towson, MD 21252, USA}
\altaffiltext{1}{Now at NASA-Goddard Space Flight Center, Greenbelt, MD 20771.\email{joshua.b.tyler@nasa.gov}}}

\begin{abstract} 
The diffuse cosmic background radiation in the GALEX far ultraviolet (FUV, 1300~\AA\ - 1700 \AA) is deduced to originate only partially  in the dust-scattered radiation of FUV-emitting stars:  the source of a substantial fraction of the FUV background radiation remains a mystery.  The radiation is remarkably uniform at both far northern and far southern Galactic latitudes, and it increases toward lower Galactic latitudes at {\em all} Galactic longitudes.  We examine  speculation that it might be due to interaction of the dark matter with the nuclei of the interstellar medium but  we are unable to point to a plausible mechanism for an effective interaction.  We also explore the possibility that we are seeing radiation from bright FUV-emitting stars scattering from a ``second population" of interstellar grains---grains that are small compared with FUV wavelengths.  Such grains are known to exist (Draine 2011) and they scatter with very high albedo, with an isotropic scattering pattern.  However, comparison with the observed distribution (deduced from their $100\ \mu$m emission) of grains at high Galactic latitudes shows no correlation between the grains' location and the observed FUV emission.  Our modeling of the FUV scattering by small grains also shows that there must be remarkably few such ``smaller" grains at high Galactic latitudes, both North and South; this likely means simply that there is very little interstellar dust of any kind at the Galactic poles, in agreement with Perry \& Johnston (1982).  We also review our limited knowledge of the cosmic diffuse background at ultraviolet wavelengths shortward of Lyman $\alpha$---it could be that our ``second component" of the diffuse far-ultraviolet background persists shortward of the Lyman limit, and is the cause of the re-ionization of the Universe (Kollmeier et al. 2014).
\end{abstract}

\keywords{dust, extinction---ISM:  clouds---ultraviolet: ISM---Dark matter}

\section{Introduction}

Diffuse celestial background radiation is observed over  every wavelength range.  In the microwave the results of observation have been sufficiently important that they have led to physics Nobel prizes for the observers.  What about the cosmic background in the ultraviolet?  Two reviews of the ultraviolet observations (Bowyer 1991, and Henry 1991) came to apparently quite different conclusions concerning the origin of the observed (at that time) diffuse ultraviolet emissions.  Bowyer concluded that the diffuse emission includes dust-scattered starlight; we will see that Bowyer was correct.  Henry discounted dust-scattered starlight and concluded that the observed background at the highest galactic latitudes must have an exotic origin.  Henry, too, was correct, as we shall demonstrate.  The importance of the study of the UV background is brought out by Murthy (2009).

Today, a powerful new diffuse UV background dataset is available, thanks to the GALEX mission (Martin et al. 2005)---these new data have been presented by Murthy, Henry, \& Sujatha (2010), and further analysis of that data set (in particular, the ``Total FUV" of column 3 of their Table 1) will form the focus of the present paper.  

The GALEX diffuse background has also been discussed recently by Hamden, Schiminovich, \& Seibert (2013), who are able to fit most of the FUV background with what might be expected from FUV starlight scattered from dust.  We will, in the present paper, exhibit a simple radiative-transport model of dust scattering which strongly supports the Hamden et al.  fit in many respects.  They also draw attention to ``a $\sim 300$ continuum unit FUV isotropic offset which is likely due to a combination of air glow (likely the dominant contributor), a small extragalactic background component including continuum light from unresolved galaxies, and/or a Galactic component not traced by other indicators."  We will find that it is their last-mentioned possibility that best accounts for most of that signal.  This ``exotic" component is explored in the present paper---we will see that it is not simply an isotropic offset, it is a strong ``second component" of the diffuse FUV background, of unknown (but Galactic) origin.

Separating these two dominant sources of diffuse far-ultraviolet emission---the dust scattered starlight (rather mundane); the second component (exotic)---is particularly difficult because both sources have very similar spectral distributions.  Dust-scattered starlight has a flat spectrum, simply because the source spectrum (hot stars) is flat, and the albedo and scattering pattern of the interstellar grains are close to being wavelength-independent.  But the moderate-to-high-Galactic-latitude emission (which we will be showing, below, to be dominated by something  {\em other than} dust-scattered starlight) {\em also} has a flat spectrum, as measured by Anderson et al. (1979; $285 \pm 32$ photons cm$^{-2}$ s$^{-1}$ sr$^{-1}$ \AA $^{-1}$ over 1250 \AA\ - 1700~\AA) and by Tennyson et al. (1988; $300 \pm 100$ photons cm$^{-2}$ s$^{-1}$ sr$^{-1}$ \AA $^{-1}$ over 1750 \AA\ - 2800~\AA). The spectral resolution, in both cases, was about 60 \AA. This spectral similarity puts the onus on the present authors to demonstrate that this ``second component" of the diffuse FUV background is truly (mostly at least) {\em not} merely dust-scattered starlight---and attempting to establish that, is the principal aim of the present paper.

\section{Overview of the data}

We focus on analysis and discussion of the diffuse ultraviolet background as observed with the GALEX mission's FUV imager (1350 \AA\ - 1750 \AA), for which the images are completely free of zodiacal light---which is {\em not} the case for the GALEX mission's NUV images (1750 \AA\ - 2800 \AA).   GALEX is fully described by Martin et al. (2005).

As indicated above, we will put forward in the present paper the GALEX FUV observations without allowance for ``airglow," despite certainty that a small amount of non-astrophysical signal is present.  The allowance for that, that resulted in Column 4 of Murthy, Henry, \& Sujatha 2010), is obsolete---to be replaced by Murthy (2014a)---but airglow is not a problem at the minimum-brightness locations, and is only a small fraction of the total signal elsewhere:  we can, and we will, ignore it for the present paper.

In Figure~1 we display, in an Aitoff equal-area projection of the entire sky, the average brightnesses of all of the GALEX FUV imager backgrounds as determined by Murthy, Henry, \& Sujatha (2010).  The brightnesses vary from a low of 285 photon units (which is in good accord with the result of Anderson et al. 1979), to a high of 8962 photon units close to  some very-FUV-bright stars.  The scale above the map in the figure gives the surface brightness of the observed diffuse radiation in photons cm$^{-2}$ s$^{-1}$ sr$^{-1}$ \AA $^{-1}$, or ``photon units" (which Hamden et al. call ``continuum units," and which we will sometimes call simply ``units").  The virtue of this choice of units is expounded by Henry (1999).  (Regions at the lowest Galactic latitudes---the white areas in Figure~1---most unfortunately, were not observed using the GALEX FUV detector.)

The hypothesis that we will be testing in this paper (and that we will find that we must reject) is that the {\em only} (or, at least the greatly predominant) source of this GALEX observed diffuse FUV background radiation is ultraviolet light from {\em stars} (their light being scattered by the interstellar dust); so, to begin the testing of this specific hypothesis, Figure~1 also includes (blue filled circles and dots) the observed direct FUV emission from each of the stars in our adaptation (Murthy \& Henry 1995) of the TD1 catalog (Thompson et al. 1978) --- these are the putative {\em sources} for the {\em diffuse} FUV background.  The radius of each circle is proportional to the square root of the flux from that star.

To obtain a fully-revealing overview of the observations requires a second, and complementary, plot:  Figure~2 is exactly the same as Figure~1, but instead of being centered on the Galactic center, this plot is centered on the Galactic anticenter.  Comparison of the two figures gives a better impression of the gross distribution of the cosmic diffuse ultraviolet background radiation over the sky.  A few low-Galactic-latitude stellar constellations are marked, in both figures, to allow easy orientation.

In looking at each of our figures showing the distribution of the diffuse FUV background, it is important to keep in mind that interstellar dust {\em strongly absorbs} far-ultraviolet light (whatever the origins of that light).  So we can only see FUV radiation from $\sim 600$ parsecs, at most (Hurwitz et al. 1991), at low Galactic latitudes.  Only at higher Galactic latitudes (above, say, 60$^\circ$) do we see the total diffuse FUV background with no significant absorption.

The ultraviolet-bright stars (our initially-hypothesized source for all of this light) are strongly concentrated in Gould's belt, which is tipped somewhat with respect to the Galactic plane---that inclination is quite apparent in the two figures, as the {\em brightest} diffuse emission (blue, green, in the figures) adheres closely to Gould's belt.  

There is certainly no doubt that at least {\em some} of the diffuse radiation that is observed (e.g., near Spica) is dust-scattered starlight, as discussed by  Murthy \& Henry (2011). The question is, does the radiation that is observed at other locations have that same origin, or does a significant portion of it have a separate and independent origin?  Let us begin with the question as to whether the {\em highest Galactic latitude} radiation is even astrophysical in its origin.

\section{Testing data integrity}

The diffuse background that is seen at the highest Galactic latitudes (that is, the red areas in Figures~1 and 2), although detected with very high signal-to-noise, is very faint.  Is that radiation astrophysical in its origin, or might it be of solar-system or terrestrial upper-atmospheric origin?

One important but simple test (Figure~3) is to look at the brightness of that radiation as a function of time over the more than five year history of the GALEX FUV observations.  The figure includes a dashed line at a constant level of 300 photon units.  The observations span a substantial fraction of a solar cycle.  The lower bound of the observations remains steady as a rock, indicating that solar activity cannot be influencing, directly or indirectly, what is observed.

The fact that the lowest-level GALEX FUV brightnesses that are observed agree so precisely with the diffuse background's spectrum that was reported by Anderson et al. (1979) is also encouraging, but of course the Anderson et al. observation, too, was made not far above the Earth's atmosphere.  It would be good to have observations that were made much farther from Earth; and fortunately, such observations exist:  two scans made from Dynamics Explorer (Fix, Craven, \& Frank 1989).  DE was in a polar orbit, with an apogee of 23,250 km.  Figure~4 gives these DE observations in an Aitoff projection identical to that of Figure~1.  The color scheme used, however, differs dramatically from that of Figure~1, for two reasons.  First, with regard to the DE data themselves, we have been at pains to not ``adjust" those data in any way:  what is plotted is the brightnesses straight from the Fix et al. paper (the actual numbers for these important observations were not included in their paper, so we provide them in Table 1; we thank the authors for supplying them).  Now, what does that have to do with the color scheme?  Please note on the calibration bar (at the top of the figure) that the lowest observed DE brightness is only 4 units!  Well, that value is actually 4 $\pm$ 363 units (Table 1).  The DE field of view was very small, and counting statistics were significant.  Please note the colors of the scale bar, and then note the complete lack of red or of yellow in the two DE scans of the sky!  No, the DE data agree extremely well with the GALEX data, establishing that the GALEX data are free of significant upper-atmosphere  terrestrial contamination (although {\em minor} contamination is {\em known} to be present, because of small brightness changes observed in the GALEX data as a function of time over each night-time observing period); see also Murthy (2014b).

The second change in the color scheme, is that in Figure~4 ``the blue stars are red."  We paint them red now, simply so that there is, in the figure, no overlap of color whatsoever between direct starlight, and diffuse ultraviolet light---which was not the case in Figures~1 and 2.  The present figure allows us to see Gould's belt much more clearly than in the previous figures.  In particular, we see that the overwhelming majority of FUV-bright stars are confined, not only to Gould's belt, but, in fact to that {\em half} of Gould's belt that is between Galactic longitudes $180^\circ$ and $360^\circ$ (Henry 1977).  This fact will be extremely helpful to our task of testing (and ultimately rejecting) our trial hypothesis that essentially all of the diffuse FUV radiation is simply dust-scattered starlight.


To test the degree of agreement between DE and GALEX quantitatively, we have identified all of the DE observations that were made at locations that are within each of the individual GALEX FUV observations, finding that of all of the GALEX targets, 546 were also observed with DE.  The average number of DE observations at each GALEX location was 3.27---one GALEX location was observed 9 times by DE.  We have averaged the DE observations for each of the GALEX targets to improve the statistics, and we have plotted the result, versus the GALEX observed value, in Figure~5.  The result confirms that Dynamics Explorer {\em does} detect the same background as does GALEX; indeed, the DE background is, if anything, slightly brighter than the GALEX backgroud.

What does this close agreement tell us about the GALEX observations?  Figure~6 is a cartoon showing the Earth, with the GALEX and Dynamics Explorer orbits drawn as if they were coplanar and located in the page.  The great majority of the DE diffuse background measurements were clearly made from locations that were {\em much farther}  from the Earth than were the GALEX observations.  The fact that the DE FUV background and the GALEX FUV background closely agree, as we have seen in Figure~5 that they do, gives us some confidence that the GALEX FUV brightnesses can be trusted, at the highest Galactic latitudes, to be largely astrophysical in their origin.

\section{Displaying the high-Galactic-latitude cosmic background}
 
Our focus, at least initially, will be on testing our understanding of the diffuse ultraviolet background at the {\em highest} Galactic latitudes, where the influence of Galactic ultraviolet starlight should be least.  So, rather than an Aitoff projection, we switch to separate polar projections (north Galactic, and south Galactic).  In Figure~7 we display the FUV radiation that is observed in the northern Galactic hemisphere, on a logarithmic scale similar to that of Figure~1.
 
 In these polar plots we also include, black open circles (and black dots), the TD1 FUV stars for that hemisphere, as well as the very brightest FUV stars of the other hemisphere ({\em dashed} open circles).  White areas again are regions with no GALEX FUV observations; many of these contain black circles or black dots revealing {\em why} they were not observed by GALEX:  they are the locations of FUV-bright stars (which might have damaged the GALEX detectors).
 
 Our earlier plots of the observations were for orientation; this plot, and following plots, will be for analysis and critical discussion.  The lack of color in the star plotting symbols avoids any possible confusion between direct starlight and the diffuse emission.  This plot also brings out the remarkable confinement of the brightest FUV stars to regions near the galactic plane; that is, the rim of this figure (though keep in mind that the projection used emphasizes display of the highest latitudes), and specifically to the {\em top }rim of this figure (and of following figures).  
 
 The brightest northern-hemisphere individual stars that were seen in Figures 1 and 2 repay inspection:  only Spica provides good evidence for an origin of at least some of the {\em broader} observed diffuse radiation in dust-scattered starlight.
 
 Note that the brightest stars are, as we have already seen, not only near the Galactic plane, but also overwhelmingly confined to the longitude range 180$^{\circ}$ to 360$^{\circ}$ (which is the top half of the plot).  Notice, very importantly, in contrast, that there is no asymmetry at all in the diffuse background over most of this plot.  This, by itself, argues that the diffuse background at the higher Galactic latitudes can hardly originate in dust-scattered starlight.
 
 An important complement to Figure~7 is the corresponding {\em linear} intensity plot, Figure~8.  The great virtue of the linear plot is that its lowest brightness is zero.  (On the other extreme of our new intensity scale, the brightnesses are cut off at 2000 units; regions brighter than that value are shown as white).
 
 We will be using these plots to test (and, we will see, to reject) the hypothesis that what we are seeing in the figures is exclusively (or even predominantly) dust-scattered starlight.  We are fortunate that {\em both} celestial Galactic hemispheres are observed by GALEX.  Figures 9 and 10 are the same as the previous two figures, but this time for the opposite, southern, Galactic hemisphere.  Again the Galactic longitude range to which the brightest stars of Gould's belt are confined, is over the top half of the figure.  One or two southern Galactic hemisphere stars show evidence for some of the diffuse background being dust-scattered starlight; these, of course, are among the stars discussed by Murthy \& Henry (2011), that confirm a very strongly {\em forward} scattering property, for FUV radiation, of the interstellar dust.
 
 While forward-scattered light is easy to detect because of its concentration around the location of the source star, Draine (2011) points out that ``the tendency for the extinction to rise with decreasing $\lambda$, even at the shortest wavelengths where we can measure it, tells us that grains smaller than the wavelength must be making an appreciable contribution to the extinction, down to $\lambda = 0.1\ \mu$m."  So we must be sensitive to the fact that some FUV radiation will be scattered, not forward, but isotropically.
 
 The two methods of display that we have used, linear and logarithmic, are both of value.  The logarithmic brings out clearly the variations in brightness that occur from place to place over the polar caps, while the linear shows that these variations are on top of a more uniform base emission that is present everywhere.  
 
 We are testing the hypothesis that the diffuse radiation that is mapped in Figures ~7 and~8, and~9 and~10, originates, at least mostly, in starlight scattered from interstellar dust, and our tentative rejection of that hypothesis, so far, rides largely on the {\em uniformity} of the faintest diffuse radiation, in both hemispheres, compared with the strong non-uniformity in the distribution of the radiation sources (stars).  But the radiation is not completely uniform; it clearly increases in brightness (yellow regions) toward lower Galactic latitudes, at all Galactic longitudes.  It is striking that that increase is almost entirely independent of Galactic longitude:  if we were seeing starlight scattered from interstellar dust, surely there would be a very strong top-half-of-figure, bottom-half-of-figure, asymmetry?  But no hint of such an asymmetry can be seen.
 
 This provides us with critical information about our putative exotic ``second component" of the diffuse far-ultraviolet background, namely that at least a portion of it {\em increases toward lower Galactic latitudes}.  That argues conclusively that whatever the second component may be, at least some portion of it is Galactic, not extragalactic, in its origin; and, equally conclusively, that at least that portion is celestial, and not terrestrial, in its origin.
 
 The preceding four figures showed the distribution over both Galactic hemispheres of the GALEX FUV diffuse background radiation.  To avoid confusion, in  these figures direct starlight was only indicated using black circles and black dots.  In Figure 11, we provide a ``finder chart" for the FUV-bright stars of both hemispheres:  blue for northern hemisphere stars, and red for southern hemisphere stars.  As in some previous figures, constellation names are provided around the rim of the figure (the Galactic plane), and, this time, star names are provided for some of the brightest stars.  Once again we note the extraordinary confinement of the brightest UV-bright stars to the lowest Galactic latitudes:  the three circles are at Galactic latitudes 0$^\circ$, 30$^\circ$, and 60$^\circ$.

\section{Comparison with the $100\ \mu$m thermal emission}

In Figures 7-10 we have displayed the diffuse ultraviolet background, with our view being centered on the Galactic poles.  The hypothesis that we are testing is that this radiation is (largely, at least) ultraviolet starlight that has been scattered from interstellar dust.  We would of course like to compare the distribution of the ultraviolet emission, with the distribution of the dust that is supposedly doing the scattering.  Where is that dust located?  Fortunately, we can locate the dust with no ambiguity (that is, as far as its angular distribution on the sky is concerned; we have no idea of its distance)---for the dust is not cold, it is somewhat heated by starlight, and therefore it emits infrared radiation.  And so, maps of the $100\ \mu$m thermal emission will show us where the dust is located on the sky.

Figures 12 and 13 display, on a logarithmic intensity scale, the northern and southern Galactic hemisphere distributions of the $100\ \mu$m cosmic thermal emission (Schlegel, Finkbeiner, \& Davis 1998).  More recent Planck observations (Abergal et al. 2014) show that the dust density varies on scales smaller than sampled by Schlegel et al., but are in general agreement.  New measurements, with Pan-Starrs1 (Schafly et al. 2014), are also in good general agreement with the Schlegel et al. measurements.

We will shortly be making a  {\em quantitative} comparison of the FUV radiation with this $100\ \mu$m emission; but even qualitative examination of the distributions is highly instructive.  In particular, a highlight of the polar distributions of the FUV emission, we saw, was its {\em uniformity} over both polar caps.  But now, our two figures displaying the distribution of the thermal emission (that is, displaying the distribution of the interstellar dust) reveal that the dust is anything-but-uniformly distributed!

Of course we are assuming that we know with certainty that the distribution of the $100\ \mu$m emission gives us the distribution of the interstellar dust.  This is not at all a controversial assumption, but it is still interesting to examine and test that hypothesis too!  And the present figures give an excellent means of scrutinizing that claim.  The interstellar dust is heated, and the interstellar grains then radiate the infrared radiation.  But what exactly is it that heats the interstellar dust?  Why it is, largely, the very same FUV-bright stars that are, in the hypothesis that we are testing, generating the FUV background that is the subject of the present paper!  And we have already noted that the great majority of those stars are located around the upper half of the rim of Figures~7-13.  But notice, now, that the asymmetry in the $100\ \mu$m emissions that we have already noted in the two hemispheres is {\em a quite different} asymmetry in each of the two hemispheres!  In the Northern hemisphere, the  faintest $100\ \mu$m emission is located  farthest from the top rim of the figure, where the great majority of the FUV originates.  Is that faintness due to angular distance from the heating stars; or is it due simply to lack of interstellar dust?  To answer that question, simply glance at the  southern Galactic hemisphere $100\ \mu$m emission!  For this hemisphere, the asymmetry is almost exactly opposite to what we have just noted in the northern Galactic hemisphere.  The dust that is closest to the heating source is the faintest!  Our test confirms the conventional view that the $100\ \mu$m emission plots simply show us the (very non-uniform)  location of the interstellar dust:  which is exactly what we want to know.

All of this is qualitative, but it does give us confidence that our basic understanding of the $100\ \mu$m emission's origin is correct: the infrared emission is showing us, accurately, the (angular)  location of the interstellar dust.  If the $100\ \mu$m emission is faint, that means there is little dust in that direction:  and so, if the FUV emission is relatively {\em strong} at such locations, well, the FUV emission (at least mostly) cannot be FUV starlight that has been scattered from interstellar dust. And so, that {\em is} our conclusion, from this simple, qualitative, comparison of the FUV and infrared emissions from the two hemispheres.  

We next  greatly strengthen our conclusion by a {\em quantitative} analysis of these same data sets.

\section{Quantitative comparison with the $100\ \mu$m thermal emission}
 
Glancing back at our previous figures, we see that the FUV brightnesses were plotted in units of photons cm$^{-2}$ s$^{-1}$ sr$^{-1}$ \AA $^{-1}$, while the $100\ \mu$m emission was in units of MJy sr$^{-1}$.  We now want to plot these observations together, in a single plot, so as to compare their relative strengths, as a function of Galactic latitude.  We must therefore use the same units for the brightnesses in the two distinct wavelength ranges.  What units should we use?  In deciding what set of units to use, we need to keep in mind what our purpose is.  Energy is conserved. The hypothesis that we are testing is that ultraviolet energy from the hot stars of (mostly) the Galactic plane, encounters the interstellar dust at high northern and southern Galactic latitudes, and some of it is simply scattered and provides the diffuse FUV background that we detect with GALEX; while some of it is absorbed by the interstellar grains, and then is re-emitted isotropically, as infrared radiation, which we observe at $100\ \mu$m.
 
 This reexamination of our purposes informs us as to what units we should use:  Henry (1999) has shown that units of photons cm$^{-2}$ s$^{-1}$ sr$^{-1}$ \AA $^{-1}$ are the proper choice for a spectral plot when comparisons of energy content is the purpose of that plot---as it is, in the present case.
 
 So!  We must convert our infrared brightnesses from MJy sr$^{-1}$ to photons cm$^{-2}$ s$^{-1}$ sr$^{-1}$ \AA $^{-1}$.  That is, given $n$, we need to find $x$, in the equation 

\begin{equation}
n\mbox{ MJy sr}^{-1} = x\mbox{ photons cm}^{-2} \mbox{s}^{-1} \mbox{sr}^{-1} \mbox{\AA} ^{-1}
\end{equation}

But 1 Jy = 10$^{-23}$ erg cm$^{-2}$ s$^{-1}$ Hz$^{-1}$ is the definition of the Jansky (Jy), so multiplying both sides of this Jansky-definition equation by 10$^{6}$ (as well as by $n$), and also by sr$^{-1}$, we write

\begin{equation}
n\mbox{ MJy sr}^{-1} =n\mbox{  10} ^{-17}\mbox{  erg cm}^{-2} \mbox{s}^{-1} \mbox{Hz}^{-1} \mbox{sr} ^{-1}= x\mbox{ photons cm}^{-2} \mbox{s}^{-1} \mbox{\AA}^{-1} \mbox{sr} ^{-1}
\end{equation}

Two conversions remain:  from ergs to photons and (somewhat trickier) from Hz$^{-1}$ to \AA $^{-1}$.

First,

\begin{equation}
E=h\nu=\frac{hc}{10^{-8}\lambda_{\AA}}\mbox{   erg photon}^{-1}
\end{equation}

Dividing the left hand side of Equation (2) by this, we have

\begin{equation}
n\mbox{  10} ^{-17}\mbox{ erg} \;    \;   \frac{1}{        \frac{hc }{10^{-8}\lambda_{\AA} }      }  \;           \frac{1}{\mbox{erg photon}^{-1}}                       \mbox{ cm}^{-2} \mbox{s}^{-1} \mbox{Hz}^{-1} \mbox{sr} ^{-1}= x\mbox{ photons cm}^{-2} \mbox{s}^{-1} \mbox{\AA}^{-1} \mbox{sr} ^{-1}
\end{equation}

\noindent{or}

\begin{equation}
n\mbox{  10} ^{-25} \; \,    \frac{\lambda_ {\AA} } {hc}                 \,\,                \mbox{photons cm}^{-2} \mbox{s}^{-1} \mbox{Hz}^{-1} \mbox{sr} ^{-1}= x\mbox{ photons cm}^{-2} \mbox{s}^{-1} \mbox{\AA}^{-1} \mbox{sr} ^{-1}
\end{equation}

The bandwidth conversion is less mechanical.  First, $\nu=\frac{c}{\lambda}$, so $\Delta \nu=\frac{-c}{\lambda^2}   \Delta \lambda\  $ and

\begin{equation}
\Delta \nu\;_{Hz}=\frac{-c}{\lambda_{cm}\lambda_{\AA}} \;\;\Delta \lambda\ _{\AA}
\end{equation}

The minus sign merely recognizes the fact that as wavelength increases, frequency decreases; it may be omitted.  Inserting this, our final conversion, gives us

\begin{equation}
\frac{c}{\lambda_{cm}\lambda_{\AA}}\; \mbox{Hz \AA}^{-1}   \:   n\mbox{  10} ^{-25} \; \,    \frac{\lambda_ {\AA} } {hc}                 \,\;                \mbox{photons cm}^{-2} \mbox{s}^{-1} \mbox{Hz}^{-1} \mbox{sr} ^{-1}= x\mbox{ photons cm}^{-2} \mbox{s}^{-1} \mbox{\AA}^{-1} \mbox{sr} ^{-1}
\end{equation}

\noindent or

\begin{equation}
x=n\times \frac{10^{-25}}{h\lambda_{cm}}
\end{equation}

\noindent which is the transformation that we want.  (We hope that this tutorial on the transformation of units is useful to beginning graduate students;  possibly to some others as well.)

We now apply this transformation to the case at hand:  $n=1$ MJy sr$^{-1}$ corresponds to

\begin{equation}
\frac{10^{-25}}{6.625\times 10^{-27}   \times100\ \mu m  \times10^{-4}  \mbox{ cm }  \mu m^{-1}}=x=1509.4\mbox{ units}
\end{equation}

\noindent which provides the correspondence that we were seeking. Particularizing to our lowest observed FUV background,
 
 \begin{equation}
300 \mbox{ photons cm}^{-2} \mbox{s}^{-1} \mbox{\AA}^{-1} \mbox{sr} ^{-1} = 0.2 \mbox{ M Jy sr}^{-1}
\end{equation}

We are now in a position to make our quantitative comparison---which we do in Figure~14, where we have plotted (black) the diffuse FUV background as a function of Galactic latitude, both north and south.  But our plot includes, also, the $100\ \mu$m background, which---having learned ``what makes sense," from examination of Figures 12 and 13---we have here segregated by color:  the $100\ \mu$m background for the Galactic longitude range 30$^\circ$ to 210$^\circ$ is plotted in red, while the same, for  the Galactic longitude range 210$^\circ$ to 0$^\circ$ to 30$^\circ$, is plotted in blue.  Note the drastic difference in the ratio of red to blue at northern Galactic latitudes, compared with the ratio of red to blue at southern Galactic latitudes.  (In Figure 15, we verify that no such segregation occurs for the FUV background radiation.)

Figure 14 is the critical plot that allows us to conclude with confidence that the diffuse FUV radiation that we see at each polar cap is {\em not} starlight scattered from interstellar dust: for in this plot we see that there is far too much of it, and it is very wrongly distributed.

Perhaps at one's first glance at Figure~14, one might feel that (ignoring the more important longitude-distribution problem for a moment) possibly one {\em can} explain it all as originating in the FUV light of Galactic plane stars:  there is perhaps 2 to 4 times as much thermal radiation as there is FUV radiation, so we might think: most FUV radiation is absorbed (and then re-radiated as infrared) while perhaps a third or less is simply scattered.  But that will not do, for while the thermal radiation is emitted {\em isotropically}, at least a substantial fraction of the FUV radiation is very strongly forward-scattered (see Spica).  For us to detect the scattered light at high Galactic latitudes requires that it be scattered at more than 90$^\circ$.  But at most only some small fraction of the scattered light can be other than almost-directly forward scattered.  So, especially given the more important longitude-distribution problem, this figure seems to rule out any possibility that the diffuse FUV background at high Galactic latitudes originates significantly in starlight scattered from forward-scattering interstellar grains.  

Thus our conclusion is that  we must reject the hypothesis which we posed for this paper.  The diffuse FUV background at high Galactic latitudes (and increasing in brightness toward lower Galactic latitudes) is, predominantly, {\em not} starlight scattered from interstellar dust.  Could more sophisticated dust models explain the observations?  In Section 11 we will consider the possibility that there is a substantial amount of much smaller, {\em isotropically}-scattering interstellar grains; but, there, we will find that that does not explain the observations.

\section{Extragalactic Far-Ultraviolet radiation?}

We have just shown that at least a substantial part of the FUV background originates in an unknown source in our own Galaxy.  There is still bound to be some contribution from other galaxies, and those galaxies emit with a spectrum that, if we (momentarily) ignore evolution and redshift effects, is somewhat similar to the observed {\em flat} spectrum of the diffuse light that was observed at high Galactic latitudes as measured by Anderson et al. (1979) and by Tennyson et al. (1988) using rocket-borne spectrometers.  

To estimate the size and shape of this contribution, we plot in Figure 16 two theoretical models of the spectral intensity of the extragalactic background light (EBL) at UV wavelengths, due to Finke et al. (2010, ``Model C) and Dominguez et al. (2011, upper and lower limits).  Also shown in this figure are the model predictions from a code that was originally written to compute EBL intensity at near-optical wavelengths (Overduin \& Wesson 2004). This code takes as inputs the spectral energy distributions of both quiescent and star-forming galaxies from FUV to sub-mm wavelengths (Devriendt et al. 1999).
It integrates these spectra over redshift and incorporates both galaxy number and luminosity evolution by normalizing the total luminosity density at each redshift to a mix of theoretical modeling and observational data, as compiled by Nagamine et al. (2006).
The code includes a model for absorption by dust in the intergalactic medium, due to Loeb \& Haiman (1997).
Here we apply it to calculate the spectral intensity of the EBL at FUV and NUV wavelengths, as shown in Figure~16.
There are four theoretical curves (labeled TVD, SA, Fossil and H\&S in the figure) refering to different galaxy evolution models from Nagamine et al. (2006), and the parameters of these models have been chosen to achieve the widest possible spread in model predictions.
They can probably be regarded as firm upper limits on EBL intensity at FUV wavelengths, due to the limitations of the dust opacty model and galaxy SEDs employed (Overduin, Prins, \& Strobach 2014).  All model predictions assume a standard $\Lambda$CDM cosmology with with $\Omega_{M}=0.3$ and $\Omega_{\Lambda}=0.7$.

In Figure~16, we compare these predictions with forty years of observational constraints from rocket-borne detectors (Lillie \& Witt 1976, Anderson et al. 1979, Jakobsen et al. 1984, Tennyson et al. 1988), Apollo~17 (Henry et al. 1978), Apollo-Soyuz (Paresce et al. 1980), Solrad-11 (Weller 1983), Voyager~2 (Holberg 1986), shuttle-borne spectrometers (Murthy et al. 1990, Martin et al. 1991), Dynamics Explorer-1 (Witt \& Petersohn 1994), DUVE (Korpela et al. 1998), EURD/Minisat (Edelstein et al. 2001) and HST/Las Campanas (Bernstein et al. 2002).
Filled symbols indicate mostly spectroscopic data, while open symbols indicate mostly photometric ones.

The integrated light from distant galaxies is certainly significant, and may contribute to the excess observed by GALEX at high Galactic latitudes, particularly for longer wavelengths.
However, it is most important to note that the model predictions for EBL intensity vary strongly with wavelength, while the observational data---especially those we would regard as most solid (the spectroscopic measurements denoted by filled symbols in the plot)---show a brightness that is independent of wavelength.
They are also considerably smaller in magnitude.
As we already have presented evidence for a Galactic source that decreases toward higher Galactic latitudes, we think it reasonable to conclude that what is observed at the  highest Galactic latitudes, especially at shorter wavelengths, is mostly the same source that is producing the radiation that is observed at the lower Galactic latitudes.

\section{Modeling forward-scattered FUV starlight}
 
We have created a simple single-scattering model (Figure 17) for the expected diffuse FUV background originating in starlight scattering from interstellar dust.  We describe the model below.  In Figure 18, we show the result of the model, for the important case of  Henyey-Greenstein (1941) scattering parameter $g = 0.78$ (which is strong forward scattering), and albedo $a = 0.62$, those being the values that are reported by Hamden et al. (2013) from their analysis of the GALEX data.  We have run the model for a variety of values of the albedo and, most importantly, of the scattering parameter.  We only obtain agreement with the observations if we use strong forward scattering; otherwise the model predicts far too much scattered light at high Galactic latitudes.  (In Section 11 we will find that although a small population of small, isotropically-scattering grains could easily account for the absolute amount of radiation that we see at the highest Galactic latitudes, the distribution on the sky is still quite wrong; and we will conclude from this simply that there is very little dust indeed---large grains or small---at the highest Galactic latitudes.)

Our model uses our catalog of FUV star brightnesses (Murthy \& Henry 1995) as the sole original source of all FUV radiation.  The only additional parameters in the model are the amount of dust, and the dust scale height above and below the Galactic plane, which we take to be 100 pc (use of a Gaussian instead of an exponential produces little difference).
 
The lowest brightnesses predicted by our model are 148 photon units at high Galactic latitudes.  This is undoubtedly high for these far-northern and far-southern locations, as our model assumes a uniform distribution of interstellar dust, with no allowance for the fact (Perry \& Johnston 1982) that there is, in fact, very little dust at our particular location in the Galaxy.  The highest brightness that our model predicts is 22,253 units at $\ell = 299.8, b = -1.0$, which is in the Coalsack nebula:  Sujatha, Murthy, Shalima, \& Henry (2007) reported a {\em Voyager 1} observation (at 1159 \AA) of 23,700 units at $\ell = 301.7, b = -1.7$).  
 
 What is shown in Figure 18 is the prediction of our model for the entire sky.  Figure 19 is the same model as in Figure 18, except that now we only include that portion of the sky that was actually observed by GALEX in the FUV.  For convenience in comparison with the observations,  the top line of numbers in the figure is the actual brightness scale (which is the line just below), arbitrarily multiplied by 1.706 so as to force agreement with the brightest FUV background observations (Figures 1 and 19).
 
In the creation of this model, there are three steps (see Figure 17) to the calculation (for each distance in a given direction of observation on the sky) of the contribution to the predicted background of the dust-scattered light of each star in our catalog.

1)  We first calculate the FUV flux arriving at a particular spot on the line of sight in question, taking into account the attenuation, by the interstellar dust, of the light from that particular star.  For this stage, we assume that the light that is {\em scattered} (as opposed to absorbed) is strongly {\em forward} scattered, so that it, effectively, is simply not removed from the beam.  There are two parts to this calculation:

a)  find the total optical depth $\tau$ along the path

\begin{equation}
\tau=\Sigma  \; \, \kappa \; \,ds \; \,e^{-z/h}
\end{equation}

The sum is along the path from the star to the particular spot on the line of sight; $z$ is the distance above the Galactic plane as that path is traversed; $h$ is the scale height (100~pc) that we have assumed for the interstellar medium; $\kappa$ is the absorption  coefficient, and $ds$ is the one parsec that we have used in carrying out the calculation.  For the absorption coefficient we begin with an extinction in the visible of 1.0 magnitudes, for a path length in the Galactic plane of 1 kpc, and we translate to 1500 \AA\ using the extinction curve of Cardelli et al. (1989) with $R_V=3.1$.

b) once we have calculated the total optical depth $\tau$ along the path, we calculate the flux $F_\lambda$ arriving at that particular spot on the GALEX line of sight:

\begin{equation}
F_\lambda=f_\lambda \times (D/d)^2 [\,a+(1-a)e^{-\tau}\,]
\end{equation}

\noindent where $f_\lambda$ is the brightness of the star in photons cm$^{-2}$\,s$^{-1}$ \AA$^{-1}$ $(R/D)^2$ (from our catalog of star brightnesses),  $D$ is the distance from us to the star, $d$ is the distance from the star to that particular spot (slice) on our line of sight, $R$ is the radius of the star, and $a$ is our assumed albedo of the interstellar grains.

2) Next, we calculate how much light is scattered toward us, as the light from that particular star crosses that particular slice of that particular GALEX one-degree-field-of-view line of sight.  

This is the only point in our calculation where we take into explicit account the particular value that we have chosen for the Henyey-Greenstein scattering parameter $g$, instead of just assuming ``straight-forward"  (i.e., {\em no}) scattering.  In particular,

\begin{equation}
H= \frac{1-g^2}{(1+g^2-2g \cos\theta)^{3/2}}
\end{equation}

\noindent (Henyey \& Greenstein 1941) gives the fraction scattered in direction $\theta$, where $\theta$ is the angle of deflection.

The brightness of the slice as seen by us (who are a distance $x$ away) is given by

\begin{equation}
I_\lambda=F_\lambda\frac{H}{4\pi}\frac{a}{x^2} \;   (1-e^{w\; \kappa \; e^{-\frac{x}{h}}  })
\end{equation}

\noindent where $x$ is in parsecs, $w$ is the thickness of the slice at that distance from us, and $\kappa$ is the absorption coefficient.

3) Finally, we allow for the absorption between us and that particular slice by calculating

\begin{equation}
I_{\lambda,obs}=I_\lambda \times \; [\,a + (1-a)e^{-\tau}\,] 
\end{equation}

\noindent where $\tau$ is now the optical depth from us to the slice (at distance $x$).   The optical depth is calculated in the same way that  the optical depth $\tau$ from the star to the slice was calculated.

Our predicted brightness is of course the sum of the predicted brightnesses for each slice (to infinity) on our line of sight, carried out and summed for all the stars in our catalog.

Note that for two of our steps, we assumed total forward scattering.  If we want to use smaller values of $g$ (and in particular, we are interested in the case $g=0$, corresponding to isotropic scattering---which is what is expected for {\em small} grains) we will be forced (because of the simplicity of our calculation scheme) to ignore the interaction of the starlight with the interstellar medium except at the slice itself.  That is not a terribly bad assumption, and at least gives us {\em some} idea of what we might expect to see, if indeed all that we are seeing is dust-scattered starlight.  The result for small grains will be presented in Section 11.

So, what does our model (Figures 17, 18, and 19) tell us?  Our model is to be compared with the FUV observations, which are shown in Figures 1 and 20.  To assist comparison, in Figure~21 we display a ratio of Observed to Model: (FUV Observed)/(FUV Model)$\times 0.586$, where $0.586=5253/8962$ to normalize to highest brightness, and where, for this plot, a small number of high-Galactic-latitude FUV-bright individual stars (all of the stars in Table 2 except Spica) have been arbitrarily  omitted from the model, as not contributing significantly to the {\em observed} distribution (although they do, if they are included, to the model).  First, note that our model does not do well in predicting the absolute value of the diffuse FUV background, falling short by a factor of $\sim 1.7$.  Our model for the distribution of the interstellar dust is crude, but also, of course, we have already concluded that dust-scattered starlight is only a portion of what is observed.   Also, since the Henyey-Greenstein scattering function has no physical basis, not a great deal could be hoped for from our model in terms of the detailed distribution of scattered light on the sky.  Indeed, Draine (2003) has warned that, in the ultraviolet, the use of simple Henyey-Greenstein phases functions is problematic---he suggests that discrepancies among reports of the values of these parameters in the ultraviolet may be due to reliance on these functions. So, in the present paper, we have used the  Hamden et al. proposed Henyey-Greenstein values simply as a ``straw man" for discussion.  

There are features that are present in the model of Figure 19 that are NOT present in the observations that we have displayed in the present paper.  The model shows significant asymmetry with Galactic longitude in the predicted diffuse FUV background at moderate and high Galactic latitudes, particularly southern.  We suggest that this is a robust feature regarding what might be expected if dust-scattered starlight dominates the diffuse FUV background; the complete absence of such asymmetry in the observations again confirms that dust-scattered starlight cannot be a significant source of the diffuse FUV background at moderate and high Galactic latitudes.  Even more importantly, the model predicts very low brightnesses at low Galactic latitudes at all Galactic longitudes that are far from where the bulk of the FUV-brightest stars are in Gould's belt---yet the observations show a strong brightening toward low Galactic latitudes at {\em all} Galactic longitudes.  The observations from Dynamics Explorer are particularly important in this regard.  So we can robustly conclude that even at the lowest Galactic latitudes starlight forward-scattered by interstellar dust  is not the only source of the diffuse FUV background.  There remains the possibility of a contribution to the FUV background by isotropic scattering from very small grains;  we will deal with this possibility in Section 11.

\section{Individual GALEX observations}
 
Sujatha, Murthy, Karnataki, Henry \& Bianchi (2009), and Murthy, Henry, \& Sujatha (2010), have begun the work of examining the diffuse FUV background in individual GALEX observations.  They found that the overall level of brightness at one GALEX target at moderate ($b=+38.6^\circ$) Galactic latitude could be successfully modeled with conventional sources.  However, Henry (2010) found that the model that succeeds for that observation gives incorrect results at higher Galactic latitudes, and that the model does not predict correctly the detailed observed distribution across the one-degree field of the observation.  Indeed, this observation caused Henry (2010) to abandon his long-held belief that the high-Galactic-latitude diffuse ultraviolet background was extragalactic in its origin, and the revision of that conclusion is of course strongly reinforced in the present paper.  

The result of the GALEX FUV observation of this target is crucial to supporting the interpretation of the diffuse FUV background that we are presenting in this paper.  The observation was made as part of our Guest Investigator program with GALEX.  The target was specifically proposed because (we thought at that time) the observation would support our then-held idea that the high-Galactic-latitude diffuse FUV background was extragalactic in origin.  Instead, the observation provides conclusive proof that the non-scattered-starlight component of the FUV background is Galactic in origin, and not extragalactic.  The observation is of a high-Galactic-latitude dust cloud that was discovered by Sandage (1976).  Sandage's observation appears in Figure 22, while our GALEX FUV observation is shown in Figure 23.  The GALEX exposure was 14,821 seconds, so the signal-to-noise is extremely high.   We had expected that the FUV observation would show evidence for the absorption, by the dust cloud, of our putative extragalactic source.  Instead, to our surprise (and shock, at the time) the glow is bright, and is almost perfectly uniform:  see Henry (2010) for detailed analysis.  No, what is detected is clearly forground diffuse emission of some unknown kind from the interstellar medium,  overwhelming any trace of an effect of the dust cloud behind.  This is imaging of our ``second component" of the FUV background!  

What can this second component have as its origin?  Could it arise from some kind of interaction of the dark matter with the interstellar medium?  We examine that possibility in the next section, concluding that it is difficult to support such an attribution.  Could the radiation instead be due to scattering from a population of interstellar grains that are much smaller than the FUV-wavelengths of the radiation that is being scattered (Draine 2011)?  In our final section before our conclusions, we will examine that possibility as well---and we will find that we must also reject that possibility---leaving us with our mystery.
 
\section{Hints of new physics?}

\subsection{Background radiation and dark matter}

We have presented in this paper evidence for a second component to the diffuse FUV background, beyond the starlight scattered from conventional dust.  Henry (2010, 2012) has attributed this mysterious second component to unspecified interaction of the dark matter particles with the nuclei of the interstellar medium.  We have, below, but without success, attempted to find a mechanism or mechanisms that could justify that attribution.  Our lack of success does not, of course, rule out such an origin; in particular, if the dark matter particles should turn out to be composite particles, overall electrically neutral but involving electrically charges components (think of a neutron), then perhaps our second component could originate in collisions of those dark matter particles with the nuclei of the interstellar medium.  The attraction of the idea remains  that it would account for the fact that our ``second component" is confined to the Galactic plane.  Having said that, Dvorkin et al. (2013) seem to rule out even interactions of dark matter particles possessing electric dipole moments or tiny electric charges!

In cosmology, and more recently in observations of the Galactic center, there is a rich tradition of attributing ``bumps'' and excesses of all kinds in diffuse background radiation to new physics, usually in the form of decays, annihilations or other interactions involving dark matter and/or energy (Overduin \& Wesson 2008). Neither dark matter nor dark energy need be perfectly black; and this way of searching for them is referred to as indirect detection (as opposed to the direct detection of dark-matter particles themselves).
Features from the one-time ``sub-mm excess'' in the CMB through the ``MeV bump'' and all the way out to the GZK cutoff (or apparent absence of it) in the high-energy $\gamma$-ray background have been interpreted this way.
Many such features have gone away on closer observation.
Will the Galactic UV excess found by by GALEX suffer the same fate?
We have argued above that it is robust, and that the obstacles to explaining it with conventional astrophysics (primarily dust scattering) are serious.
Before concluding, we therefore consider whether there are any natural ways to produce such an excess from the physics of the dark sector.

\subsection{Neutrinos}

One particle considered a particularly plausible dark-matter candidate throughout much of the 1990s was the massive neutrino (Sciama 1993).
Several independent lines of evidence at that time pointed to a $\tau$ neutrino of rest energy 27~eV, decaying to a lighter species plus a photon (Figure 24). Each product carried away half the rest energy of the parent, producing a signal at $\lambda\sim 900$~\AA.
One might hope that a lighter version of such a particle could be associated with a background in the neighborhood of 1200~\AA.
This background would, however, be a sharp line and not the broad ``shelf'' that we see in the data, from the Lyman limit to the far edge of the GALEX bandpass (1350-2800~\AA).
There is no hint of such a signal in observations of other galaxies and galaxy clusters where dark matter is thought to be even more concentrated than it is in the Milky Way (Overduin \& Wesson 2008).
Moreover, PLANCK measurements of the CMB, in combination with the standard picture of structure formation by gravitational instability, now imply that the sum of all three (standard-model) neutrino masses is less than 0.35~eV (Giusarma et al. 2013).
One can evade this constraint with sterile fourth-generation neutrinos, and indeed these have recently been postulated as warm dark-matter particles decaying into photons with energies as low as $\gtrsim 0.5$~keV (Abazajian et al. 2007).
However, data on structure formation now rules out warm dark matter particles in any form with masses below 3.3~keV (Viel et al. 2013).
Neutrinos do not seem to be the answer.

\subsection{Weakly-interacting massive particles (WIMPs)}
 
The leading dark-matter candidates remain weakly-interacting particles or WIMPs.
While still hypothetical, their naturalness stems from the fact that they automatically produce the right density of cold dark matter, given only the assumption of weak-scale coupling strength to matter (or vice versa).
No fine-tuning is needed; one property follows from the other, thanks to the Boltzmann equation.
The WIMP mass is also tightly constrained: it cannot be less than about 2~GeV or WIMPs will overclose the Universe.
(This condition, known as the Lee-Weinberg bound, arises because of a $m^2$-dependence in the cross-section.
Low-mass WIMPs interact so weakly that they drop out of thermal equilibrium too soon after the big bang, producing an unacceptably high relic density.)
As we will see below, this large mass makes it difficult to connect WIMPs to UV energies, and their small coupling cross-section makes it difficult to tie them to an excess as bright as that seen in the GALEX data.

WIMPs annihilate ``directly'' into photons via loop diagrams (Figure 25).  Because it involves loops, the flux of photons from this process is extremely low.
Nevertheless, it is typically considered the most promising way to discover WIMPs via ``indirect detection'' (i.e., detection via annihilation products, rather than WIMPs themselves).
This is because, while faint, these photons are essentially monoenergetic, and therefore distinct from almost any competing astrophysical background.
However, energy conservation dictates that the energy of the photon products is closely tied to that of the parent WIMPs:
$E_{\gamma}=m_{\tilde{\chi}}$ for $\tilde{\chi}\tilde{\chi}\rightarrow\gamma\gamma$ (Bergstr\"om and Ullio 1997) and 
$E_{\gamma}=m_{\tilde{\chi}}(1-m_Z^2/4m_{\tilde{\chi}})^2$ for $\tilde{\chi}\tilde{\chi}\rightarrow Z\gamma$ (Ullio and Bergstr\"om 1998).
From the Lee-Weinberg bound it follows that WIMP annihilation directly to photons can have nothing to do with the UV background.

More promising might be processes that produce secondary photons via tree-level WIMP annihilations to quarks or W/Z bosons which then ``hadronize'' and decay to charged and neutral pions, as shown schematically in Figure 26.  Example tree-level diagrams for $\tilde{\chi}\tilde{\chi}\rightarrow qq$ are shown in Figure 27.  The flux of photons produced in this way can greatly exceed that from WIMP annihilations directly into photons.
Moreover, these secondary photons have a broader range of energies, from the WIMP mass down to perhaps the pion mass scale ($\sim 100$~MeV).
This makes them harder to detect against the many competing astrophysical backgrounds, which is why they are usually ignored in indirect dark-matter searches (see however Baltz et al. 2008).
It might make them more attractive from the point of view of explaining a relatively flat spectrum (as seen by GALEX).
However, the relevant energies are still orders of magnitude beyond the FUV scale, of order $\sim 10$~eV.

WIMP scattering off nucleons in the interstellar medium is another possibility.
If the scattering is inelastic, it would knock the nucleus into an excited state, producing de-excitation photons.
This is the basis for direct detection experiments like XENON, whose $^{129}$Xe and $^{131}$Xe target isotopes have de-excitation energies in the range 10 -100~keV (Baudis et al. 2013).
Alternatively, Henry (2012) has suggested that WIMP recoils might give rise to a flux of low-energy photons by bremmstrahlung-type acceleration of charged quarks inside the nucleus.
In high-energy physics, these are called ``direct photons'' to distinguish them from the messier photons in jets.
However, fundamental limitations, both experimental and theoretical, mean that it has only been possible to study this phenomenon in the large transverse momentum regime; i.e., at energies above about 1~GeV.
It is interesting to note that there are some experimental indications of an unexplained ``soft photon'' excess beyond QED expectations at the lower end of this range (Belogianni et al. 2002).

To decide in a {\em model-independent\/} way whether WIMP scattering can be connected to the FUV excess seen by GALEX, we proceed as follows.
The largest WIMP-nucleon scattering cross-section allowed by current experiment is $\sigma = 2 \times 10^{-41}$~cm$^2$ (Aprile et al. 2012).
The lightest WIMP mass allowed by experiment is $m_{\tilde{\chi}}=9$~GeV (Agnese et al. 2013). 
Hence the largest possible number density of WIMPs is $n = \rho_{\mbox{\tiny DM}}/m_{\tilde{\chi}} = 0.03$~cm$^{-3}$, where we have used a canonical figure for the local dark matter density of $\rho_{\mbox{\tiny DM}}=0.3$~GeV~cm$^{-3}$.
Combining these numbers, we obtain a conservative upper limit on the scattering rate per WIMP of $n\sigma v = 1 \times 10^{-35}$~s$^{-1}$, where $v = 220$~km/s is the speed of WIMPs with respect to the interstellar medium.
Even if each WIMP converts its entire rest energy into 10~eV photons, the largest possible FUV ``luminosity per WIMP'' is then $1 \times 10^{-37}$~erg~s$^{-1}$.
Now consider all the WIMPs scattering off nucleons inside a spherical region whose radius corresponds to the mean free path of an FUV photon in the local interstellar medium, $R\approx 600$~pc (Hurwitz, Bowyer \& Martin 1991).
If their number density $n$ and luminosity $L$ are uniform throughout this region, then the total FUV intensity produced cannot exceed $\rho_{\mbox{\tiny DM}}n\sigma v R < 1 \times 10^{-17}$~erg~cm$^{-2}$~s$^{-1}$.
By comparison, the intensity of excess FUV radiation detected by GALEX over its bandpass (1380-2500~\AA) is $1 \times 10^{-5}$~erg~cm$^{-2}$~s$^{-1}$, assuming a flat spectrum with 300~photons~cm$^{-2}$~s$^{-1}$~sr$^{-1}$~\AA$^{-1}$.
Based on this argument it is hard to see how the excess could be connected to WIMP scattering in any form.
The cross section, which is fixed by cosmology and by the Boltzmann equation, is simply too small.

\subsection{Axions}

The second leading candidate for dark matter is the axion.
Axions are perhaps not as natural as WIMPs, in that their coupling strength does not automatically imply the correct cosmological density.
On the other hand, they require less of a leap beyond the standard model, as their existence is implied within ordinary QCD.
Moreover there are now indications that, even in the leading (supersymmetric) WIMP models, axions arise and may make up more of the dark matter than the WIMPs themselves (Baer 2013).

Axions come in two main flavors: thermal (meaning they were originally in equilibrium with standard-model particles in the early universe) and non-thermal (meaning they arose in some other way, for instance as the result of a misalignment between the initial value of the axion field and the minimum of its potential).
Most experimental attention has focused on nonthermal axions, which have masses in the $\mu$eV-meV range.
These are numerous enough to make up the required cosmological density of cold dark matter, and hard to constrain as they do not decay into standard-model particles.
(They are also known as ``invisible axions'' for this reason.)
They can however convert into photons inside magnetic fields via the Primakoff effect (Figure 28).  This is the basis for experimental efforts to detect axions from hot stellar cores using magnetic cavities (Sikivie 1983, van Bibber et al. 1989).
The expected photon flux for a 9.0~T magnetic field inside a 9.26~m cavity aimed at the Sun (as in the CERN Axion Solar Telescope or CAST) is 0.088 photons~day$^{-1}$~cm$^{-2}$~keV$^{-1}$ $(E/\mbox{keV})$ $(L/9.26\mbox{ m})^2$ $(B/9.0\mbox{ T})^2$ $\exp[-(E/\mbox{keV})/1.305]$ (Andriamonje et al. 2007).
On simple dimensional grounds one might replace such a cavity in the case of the Galactic FUV background by the ``local bubble'' of radius $\sim$600~pc, permeated by a Galactic magnetic field of mean magnitude $B \sim 0.5 \mu$G (Mao et al. 2012; note that the Sun is located about 20~pc above the Galactic plane according to Humphreys \& Larson 1995).
A similar mechanism has recently been proposed to contribute to the diffuse cosmic x-ray background and account for the unexplained soft x-ray excess in some galaxy clusters (Conlon \& Marsh 2013).
However, the corresponding flux per wavelength of $3 \times 10^7$~erg~cm$^{-2}$~s$^{-1}$~sr$^{-1}$~\AA$^{-1}$ $(\lambda/\mbox{10 \AA})^{-4.5}$ $(L/\mbox{600 pc})^2$ $(B/0.5~\mu\mbox{G})^2$ $\exp(-\mbox{10 \AA}/\lambda)$ can have nothing to do with the GALEX UV excess; it is orders of magnitude too bright, and peaks at 2~\AA\ (or 5~keV) in the X-ray band, as might be expected since these axions form in the cores of hot stars.

Thermal axions might be more promising, as they can decay directly into photon pairs, each with $E_{\gamma}=m_a/2$, via a model-dependent axion-photon coupling constant $g_{a\gamma\gamma}$ (Figure 29).
Thus thermal axions with 9~eV$ < m_a < 18$~eV might in principle be associated with a signal like that seen by GALEX.
However, any such mechanism faces considerable challenges.
As with WIMP annihilations, one difficulty would be in reconciling the essentially monoenergetic nature of these decays with the flat spectrum observed.
Calculations using the Boltzmann equation show that axions this massive would be able to provide no more than half the observed density of dark matter.
They are also strongly constrained by astrophysical considerations.
They would drain too much energy from the cores of red giant stars, disrupting helium ignition unless $m_a \lesssim 10$~eV in the simplest models (Raffelt 1996).
Upper limits on the intensity of the extragalactic background light impose a similar bound, $m_a < 8$~eV (Overduin \& Wesson 2008).
Observations in the direction of three rich clusters of galaxies tighten this limit further, to $m_a < 3$~eV (Bershady, Ressell \& Turner 1991).
It might be worth revisiting these constraints, which depend sensitively on theoretical assumptions involving the axion coupling strength.
More recently, however, an even more stringent limit has come from data on structure formation.
Axions in this mass range are light enough to act as hot dark matter.
Arguments similar to those mentioned above in connection with massive neutrinos then imply that $m_a < 2$~eV (Hannestad \& Raffelt 2004) or even $m_a < 0.4$~eV (Melchiorri, Mena \& Slosar 2007).
These are gravitational arguments, and do not depend on the details of axion-photon coupling.
Thus axions, too, fall short.

\subsection{Other candidates}

Most of the remaining dark-matter candidates from particle physics are extremely massive (by definition, more massive than the heaviest standard-model particle), and hence manifest themselves only in the high-energy $\gamma$-ray band, if at all.
Leading examples include Kaluza-Klein states (excitations of standard-model particles associated with compact extra dimensions), branons (similar states in higher-dimensional brane-world scenarios), cryptons (stable or metastable states in string theory) and WIMPzillas (heavy non-thermal relic particles).
These particles would be remoter from the FUV band than the WIMPs considered above (whose masses are tied to the masses of the W and Z bosons).
This serves to point up the essential challenge: in order to explain the GALEX excess, one needs to find some plausible connection to physics on eV scales.
Light neutrinos and axions come closest to fitting this description, but have now been decisively ruled out by arguments that are almost purely gravitational (structure formation) and therefore very hard to evade.
The other candidates we have considered here involve physics at higher energies almost ``by design,'' and it is hard to see how they could be connected to the GALEX excess without an unnatural degree of fine tuning.

Similar remarks apply to one final dark-matter candidate, the primordial black hole (PBH), though perhaps with more leeway.
PBHs are black holes that evade the upper limit on baryonic mass density because they formed before cosmic nucleosynthesis.
In principle, their cosmological density is unconstrained.
In practice, assuming they formed by gravitational instability with a standard scale-invariant spectrum of initial masses, it is possible to be quite specific about the properties they must have.
PBHs decay by Hawking evaporation at a rate $dM/dt=-\alpha/M^2$, where $\alpha \approx 7 \times 10^{25}$~g$^3$~s$^{-1}$, so those that are evaporating at the present time $t_0$ have $M_{\ast}=(3\alpha t_0)^{1/3}\sim10^{15}$~g.
The spectrum of background radiation from PBH evaporation is approximately thermal, peaking at $\lambda\sim(4\pi/c)^2 GM$.
In the standard scenario described above, this spectrum is dominated by PBHs with $M\sim M_{\ast}$, giving a sharp peak near $10^{-4}$~\AA\ (or 100~MeV).
No such line is seen in observations of the cosmic $\gamma$-ray background, leading to the conclusion that PBHs can make up at most $\sim10^{-8}$ times the critical density (Page \& Hawking 1976, Overduin \& Wesson 2008).
It is conceivable that one could evade this conclusion by imposing a low-mass cutoff on the spectrum of initial PBH masses.
They would then radiate less, and at lower energies.
In fact, the same factor of $\sim10^{-8}$ could push the peak of their contributions to the background close to the ultraviolet band.
However, it is very hard to see how such a cutoff could arise in a natural way.
Many attempts have been made to justify such a modification for other reasons, generally to connect PBHs to various phenomena, from microlensing to $\gamma$-ray bursts, but none have gained wide acceptance.
 

 \section{Small interstellar grains}
 
 The values of the albedo and Henyey-Greenstein scattering parameter $g$ that were found by Hamden et al. (2013) to account for most of the observed brightness of the FUV background radiation to their satisfaction, $a=0.62$ and $g=0.78$, are of course deduced under the assumption that the dust-scattered FUV light of stars is the predominant source of the observed diffuse FUV background---which we find in the present paper not to be the case.  From his extensive modeling,  Henry (2002) has suggested that the grain albedo might in fact be as low as  $a=0.1$.  Mathis (2002) points out that Henry's result is very uncertain (and Henry agrees), but the point is made that we do not {\em know} the value of the albedo other than from these diffuse background measurements and their attribution to scattering by interstellar dust.  To see how sensitive to the adopted albedo value the predicted background is, the reader might look at Figure~4 of Henry (2012), where a model with $g=0.58$ and $a=0.10$ is displayed---the {\em brightest} predicted diffuse FUV background is only 514 units.
 
 We have already mentioned the very important question raised by Draine's (2011) drawing attention to the isotropically-scattered light that must be produced by interstellar grains that are smaller than the wavelength of the radiation that is being scattered.  In particular, could it be that the 300 photon units background that is observed at the highest Galactic latitudes is starlight scattered from such very small grains?
 
 Our single-scattering radiative transport model (presented above) does an excellent job; so good that we even felt comfortable in attributing its failure (by a factor of 1.7) to account for the observed diffuse FUV background, to the presence of a second component in addition to that of dust-scattered starlight.  However, the excellence of our model does require that the grains be strongly forward scattering.  Despite that limitation, we will nevertheless now apply it so as to give us at least some handle on what might be expected from isotropically-scattering grains.
 
 To cope with isotropically-scattering grains, we now adapt our model to simply confine scattering to the point of intersection of our line of sight with an element of the interstellar medium at which the starlight has arrived (Figure 17; step two of Section 8).  That is, we will {\em not} make any allowance at all for scattering as the light progressed from the source star {\em to} the intersection, nor as the scattered light subsequently proceeds to the detector.  (To see how serious these omissions are, we experimentally eliminated these same items from our forward-scattering model that produced Figure 18:  the result was not dramatically different, suggesting that our result in this section can be trusted---particularly because, as we will now see, our result is dramatic indeed.)
 
 Our previous model results (Figures 18 and 19) had a non-linear dependence on the density of the interstellar dust.  But that non-linearity came entirely from the two legs of radiative transport that we have been forced to omit for the case of isotropic scattering.  So for our present test, we can scale the final brightness, if we wish, by simply increasing or decreasing the density of the interstellar dust.  Since our sole aim is to try to reproduce the uniform high-Galactic-latitude 300 photon unit background by means of scattering by small grains, we have simply adjusted the interstellar matter density as required to force a result of 300 photon units.  The factor that we find to be required is 11.045, that is, we have had to reduce the interstellar dust density by that large factor if we do wish to not  overproduce the predicted high-Galactic-latitude diffuse background!     

Our result is presented in Figure 30, and is (but only at first glance) extremely surprising.  Keep in mind the fact that our maps (Figures 9 and 10) of the $100\ \mu$m emission give the amount of thermal radiation from the dust, not the actual amount of dust. The physics that Draine presents is unexceptionable.  And Draine notes that the sharp rise in the extinction curve that occurs at the shortest FUV wavelengths certifies that small grains do indeed exist, as expected.  Yet at high Galactic latitudes, we clearly do NOT see the starlight, dust-scattered at large angles, that those facts would lead us to expect!  The answer appears to be that Perry \& Johnston (1982) were correct in asserting that there is negligible reddening within $\sim$~200~pc from the Sun on the Galactic plane.  We don't see dust-scattered starlight from small grains, simply because there is a very low density of interstellar grains of any kind, in our neck of the Galaxy.  The fact that we see little or no broadly dust-scattered light from all but one of the stars (Spica) in Table 2 is in accord with this interpretation.  Also, we do know that we are located in what is now called the ``local bubble" (McClintock, Henry, Linsky, \& Moos 1978), a sector of the Galaxy having an exceptionally low density of interstellar material (perhaps as a result of an ancient supernova explosion).  While we have emphasized the excellence of our model (at least for diffuse forward-scattered light from stars), there is one element of our model that is clearly wrong:  it assumes a completely locally-homogeneous interstellar medium.  Our model predictions for the {\em highest} Galactic latitude observations should not be trusted at all, in light of the Perry \& Johnston (1982) result; we obtain, with our model, much too high predicted brightnesses for starlight scattered from dust.

 
\section{Conclusions}
 
Very high quality observations of the spatial distribution of the diffuse FUV background, with excellent signal-to-noise, are available, thanks to the GALEX mission (Martin et al. 2005) and to the work of Murthy et al. (2010).  We have attempted to account for these observations as originating in the dust-scattered FUV light of the OB stars of our Galaxy.  We have failed in this attempt.  Which leaves us with a mystery:  there is an FUV radiation field in our Galaxy that is of unknown origin, and there seems to be no conventional source for it that is readily plausible.  Henry (2012) has been led to speculate that the radiation might originate in the particles of the dark matter of our Galaxy interacting with the nuclei of the interstellar medium.  Such interaction is perhaps possible (Baudis et al. 2013) but production of the observed flat FUV spectrum has not (or at least, has not yet) been demonstrated to be possible, and in the present paper we find ourselves unable to point to a plausible mechanism to produce such radiation.  Indeed, Dvorkin, Blum, \& Kamionkowski (2013) seem to rule out such an origin.

If indeed, as we believe to be the case, we have identified a new ultraviolet radiation field in the Galaxy, then depending on its physical origin, the possibility exists that its spectrum extends below 912 \AA; that is, that it is a source of ionizing radiation extending well above the Galactic plane.  The critical need for just such an ultraviolet source has been emphasized by Kollmeier et al. 2014, in their paper ``The Photon Underproduction Crisis." They speculate that what they call the  ``missing" ionizing photons could be coming from decaying or annihlating dark matter particles in the dense cores of halos and subhalos.  In light of their call, we draw attention to both our own observations (with Voyager) at wavelengths shortward of Lyman $\alpha$ (Murthy, Henry, \& Holberg 2012), and also to the unique J-PEX observations of Kowalski et al. (2006, 2009, 2011) at still shorter wavelengths (Figure 31).  The lack of accord that we find of the Voyager observations nearest the two J-PEX observations, with those observations, simply means that the spectrum of the shortest-wavelength background is complex.  Our Voyager observations include remarkably bright patches on the sky---we direct attention to the possibility that the observed ``overproduction" of photons that we report, in our various papers, might resolve the ``underproduction" crisis---and allow us to understand the reionization of the Universe. While that understanding would be important indeed, far more important would be the clue to new physics.

If as we suggest there is indeed a substantial component of the diffuse FUV background that is NOT simply starlight scattered from dust, confirmation would of course be highly desirable---and would not be terribly difficult to achieve.  What is needed is high-signal-to-noise spectroscopy of the diffuse FUV background at a few high-Galactic-latitude locations, to probe the detailed spectral character of the radiation.  
\acknowledgments

We thank Tom Krause for discussions on opacity and Jeff Owens for helpful comments on direct photons.
We are grateful to a host of colleagues with whom we have worked over the years.
We are grateful to Maryland Space Grant Consortium, to the  Johns Hopkins University, to Towson University, and to the Indian Institute of Astrophysics, for support.  Finally, we are grateful to a thorough and persistent referee.

\appendix

\clearpage


 \begin{deluxetable}{rrrrr}
 \tablewidth{435pt} 
\tablecaption{The Two Dynamics Explorer FUV Scans (Fix, Craven \& Frank 1989)\tablenotemark{a}}

 \tablenum{1} 
 \tablecolumns{5}
\tablehead{\colhead{Scan} & \colhead{No.} & \colhead{$\ell$} & \colhead{\emph{b}} & \colhead{Photon Units} }
 \startdata
1 &    1 &  262.56 &   36.73 &    1352 $\pm$ 533\\
1 &    2 &  262.73 &   36.54 &     717 $\pm$ 212\\
1 &    3 &  262.90 &   36.36 &     868 $\pm$ 218\\
&.......&&&\\
 1 & 1321 &  261.73 &   37.61 &     838 $\pm$ 216\\
1 & 1322 &  261.91 &   37.42 &    1231 $\pm$ 241\\
1 & 1323 &  262.08 &   37.24 &    1020 $\pm$ 241\\
---&---------&---------&---------&---------------\\

 2 &    1 &  244.50 &   45.42 &     510 $\pm$ 397\\
2 &    2 &  245.47 &   44.81 &     930 $\pm$ 427\\
2 &    3 &  245.72 &   44.65 &     340 $\pm$ 386\\
&.......&&&\\
 2 & 1282 &  242.84 &   46.41 &    1625 $\pm$ 457\\
2 & 1283 &  243.09 &   46.27 &     567 $\pm$ 393\\
2 & 1284 &  243.35 &   46.12 &     805 $\pm$ 408\\

\enddata 

\tablenotetext{a}{Table 1 is published in its entirety in the electronic edition of the Astrophysical Journal. A portion is shown here for guidance regarding its form and content.}

\end{deluxetable}

 \begin{deluxetable}{rrrrrr}
 \tablewidth{280pt} 
\tablecaption{Individual Stars}
 \tablenum{2} 
 \tablecolumns{6}
\tablehead{\colhead{\emph{z} pc}& \colhead{\emph{b} }& \colhead{Name} & \colhead{Name} & \colhead{HR} & \colhead{HD} }
 \startdata
-36.8 &    $-58^{\circ}$47 &  $\alpha$ Eri &   Achernar &    472 & 10144\\
-24.6 &    $-52^{\circ}$28 &  $\alpha$ Gru  &  Alnair &    8425  & 209952\\
-87.3 &   $-46^{\circ}$41 &  $\gamma$ Peg  &  Algenib &    39  & 886\\
-31.7 &    $-35^{\circ}$11 & $\alpha$ Pav  &   Peacock &    7790 & 193924\\
+18.3 &    $+48^{\circ}$56 &  $\alpha$ Leo  &  Regulus &    3982  & 87901\\
+62.1 &    $+50^{\circ}$52 &  $\alpha$ Vir  &   Spica &    5056  & 116658\\
+29.0 &    $+65^{\circ}$19 &  $\eta$ UMa &   Alcaid &    5191  & 120315\\
\enddata 
\end{deluxetable}


\begin{figure} 
\epsscale{1.10}
\plotone{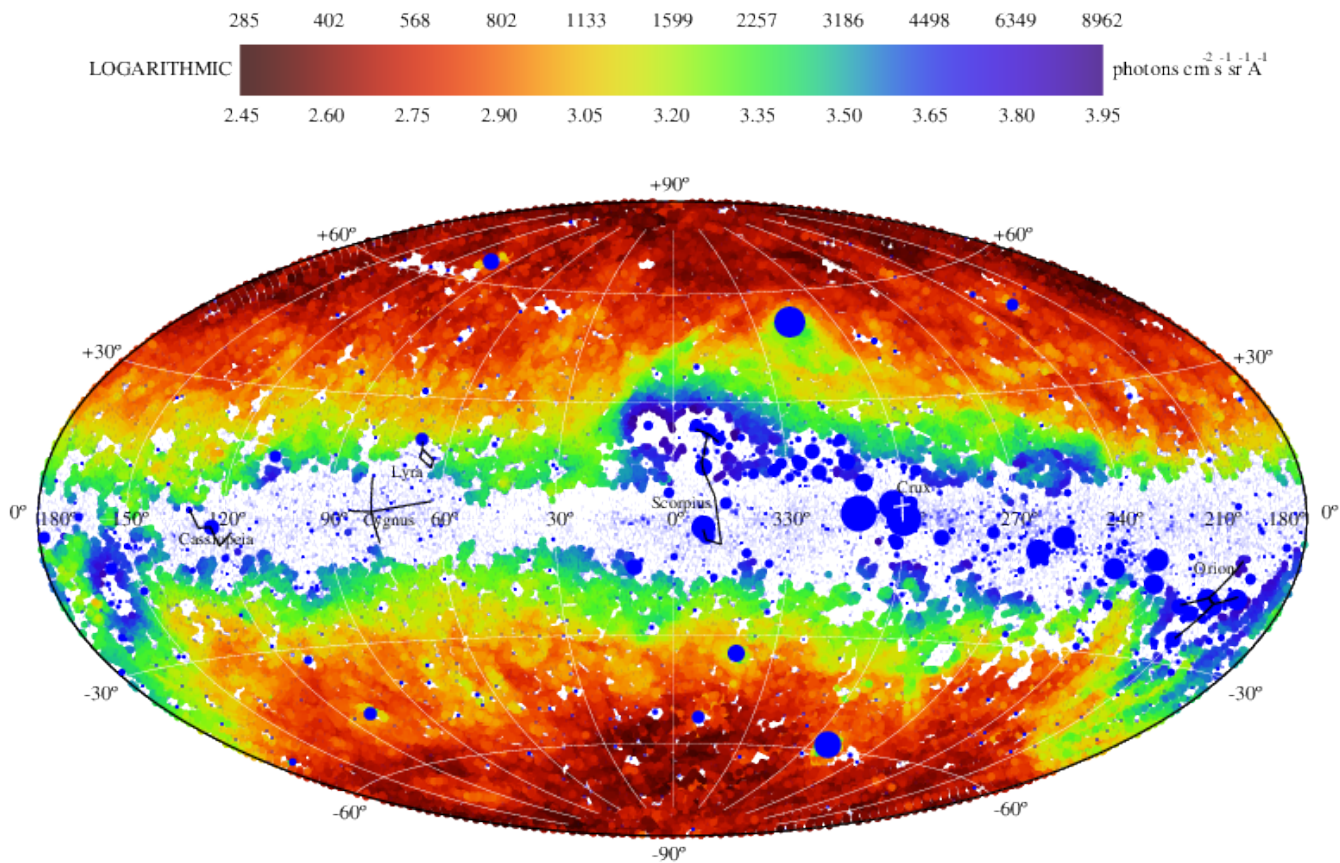}
\caption{Blue circles and dots mark the locations of 31,215 TD1 far-ultraviolet-bright stars, while the GALEX FUV diffuse background is also shown (more than 30,000 one-degree-sized spots).  (At the lowest galactic latitudes, no GALEX FUV images exist.)  A few constellations are shown for orientation purposes.  Some FUV-bright stars (e.g., in Crux) are behind significant amounts of interstellar dust, which forward-scatters their light to us (Murthy, Henry, \& Holberg 1994).  Most FUV-bright stars at high galactic latitudes have very little foreground dust; an important exception being Spica (Murthy \& Henry 2011) located at $\ell = 316^{\circ}, b = 50.8^{\circ}$.  One crucial question is, is the ubiquitous red background at high galactic latitudes astrophysical, or is it largely geocoronal?   Attempting to resolve that question is one focus of the present paper.     \label{fig1}}
\end{figure}
\clearpage

\begin{figure} 
\epsscale{1.10}
\plotone{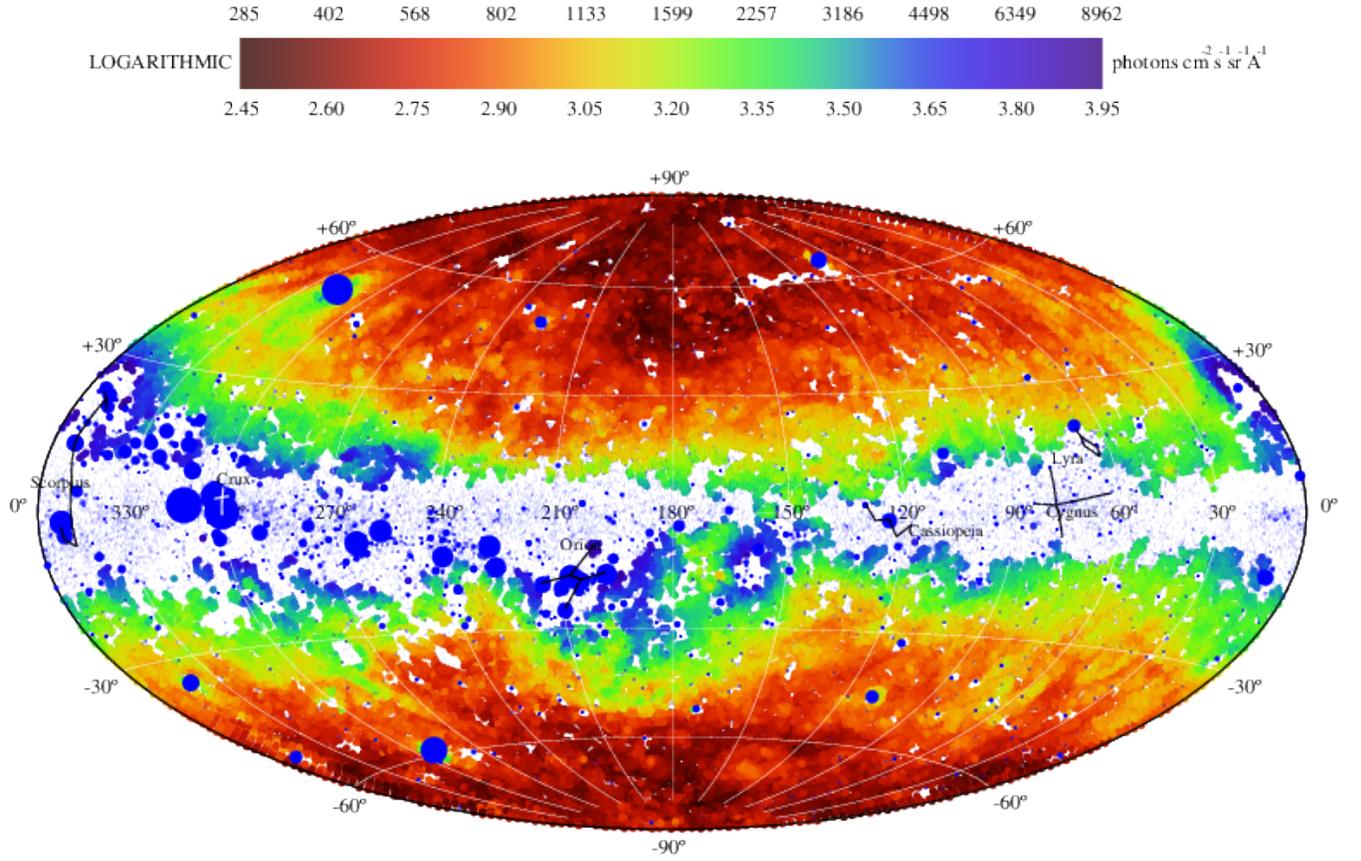}
\caption{ This is the same as Figure 1, except that it is centered on the Galactic anticenter, giving a much better display of the northern-Galactic-hemisphere distribution of the far ultraviolet background radiation.  \label{fig2}}
\end{figure}
\clearpage

\begin{figure} 
\epsscale{1.10}
\plotone{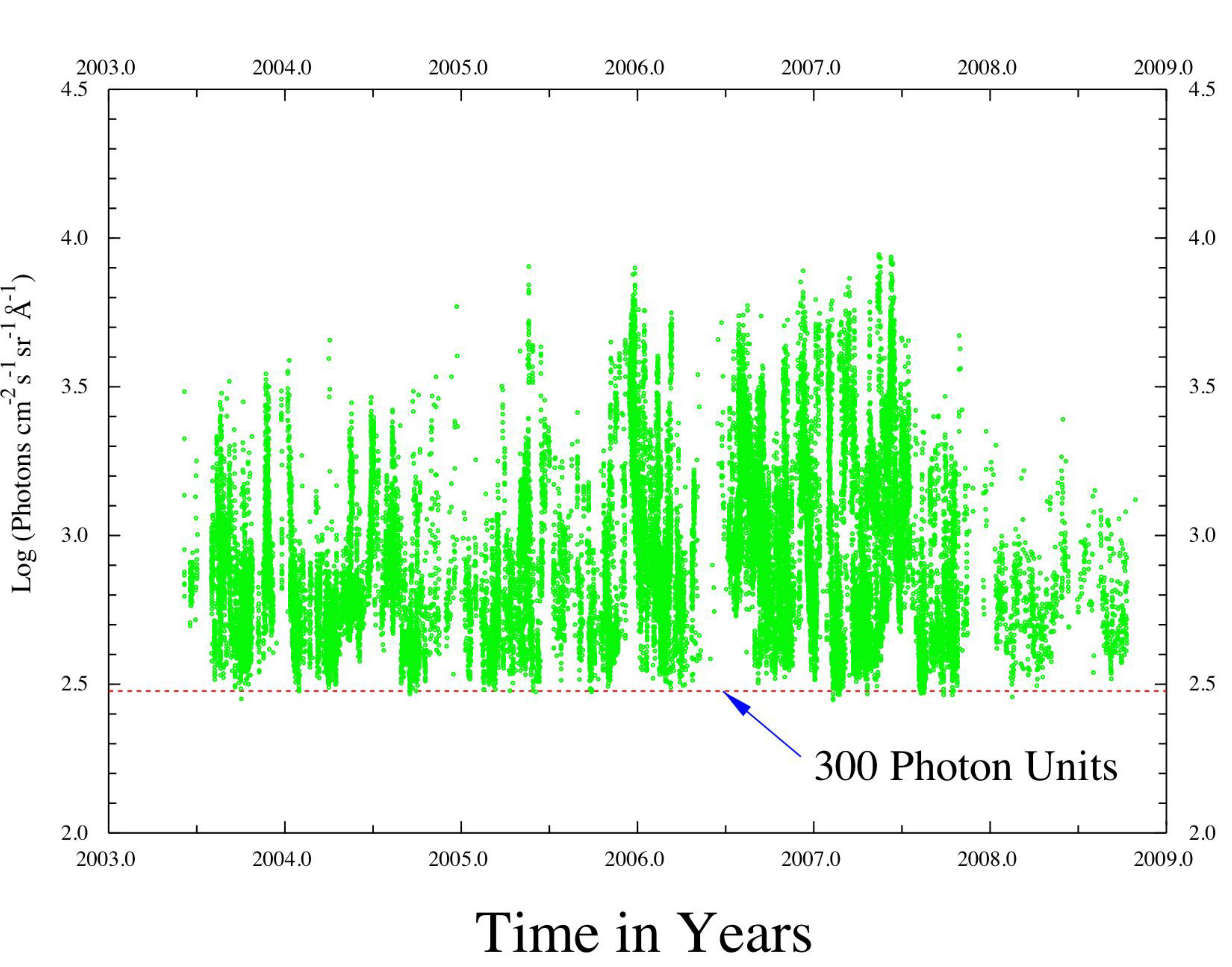}
\caption{ All of the GALEX FUV observations from the entire mission have their diffuse background measurements plotted here as a function of year of observation.  The observations were made over a substantial fraction of a solar cycle.  The minimum value observed, $\sim 300$ photon units, does not change over that span of time, strongly suggesting that the minimum FUV brightness measured by GALEX is not significantly affected by terrestrial atmospheric emission, but rather is astrophysical in its origin.  \label{fig3}}
\end{figure}
\clearpage

\begin{figure} 
\epsscale{1.10}
\plotone{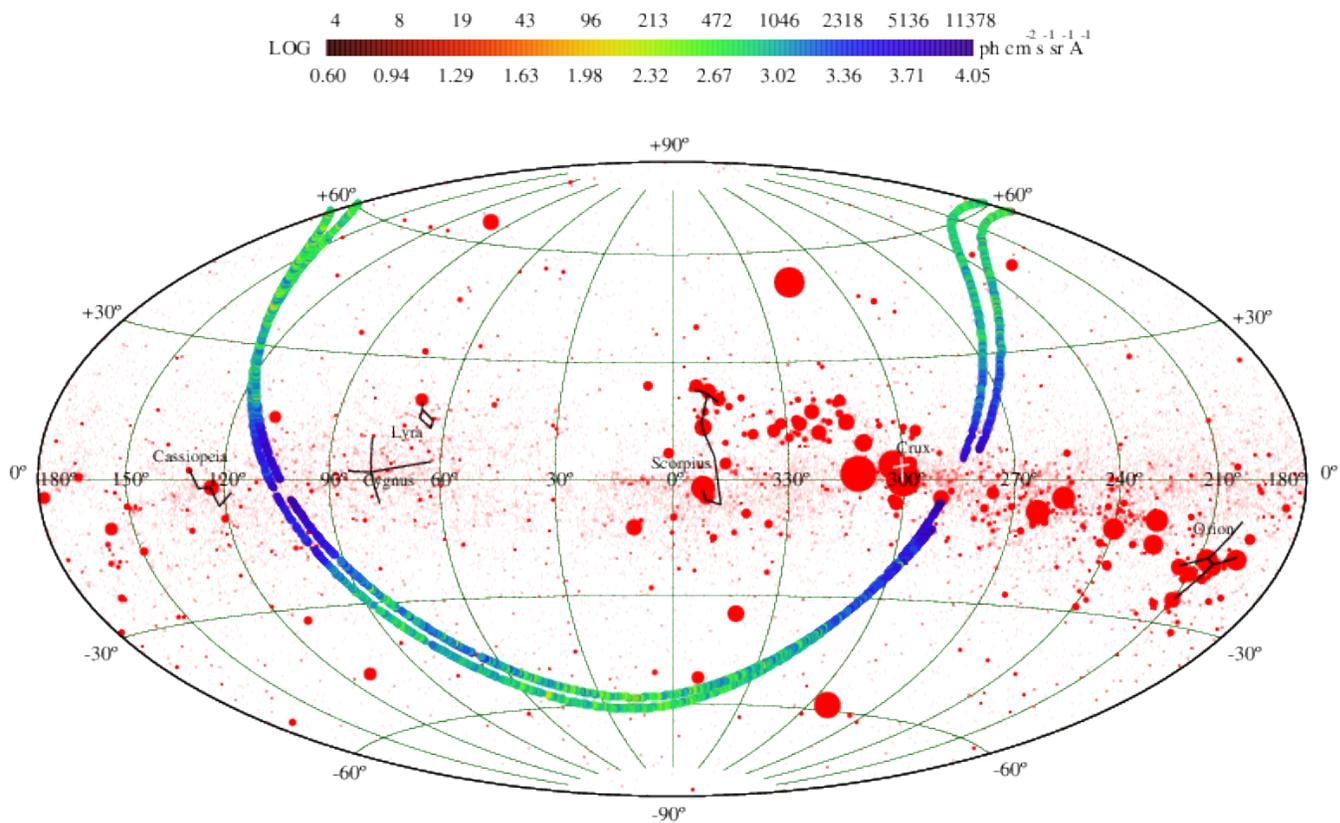}
\caption{ Dynamics Explorer (``DE", Fix, Craven, \& Frank 1989) was used to carry out two far-ultraviolet scans around the sky, with the results shown here.  (The UV-bright stars are colored red in this figure, rather than blue, simply to provide better contrast, with the DE observations, at low galactic latitudes.)  The lowest DE value was $4 \pm 363$ units, which (but, inconsequentially) distorts the color scale.  At high galactic latitudes, both DE scans detect the background at a few hundred photon units, strongly suggesting, because DE is in a much higher orbit (Fig. 6) than is GALEX, that the GALEX observations of similar values are {\em not} due to geophysical contamination.  Both longitude regions of the galactic plane that were observed by DE are seen to be very bright:  the brightest region that is seen on the scan that passes closest to Cassiopeia is $6082 \pm 455$ photon units, while the brightest region for the scan that passes slightly farther from Cassiopeia was $9783 \pm 752$ photon units.  \label{fig4}}
\end{figure}
\clearpage


\begin{figure} 
\epsscale{1.0}
\plotone{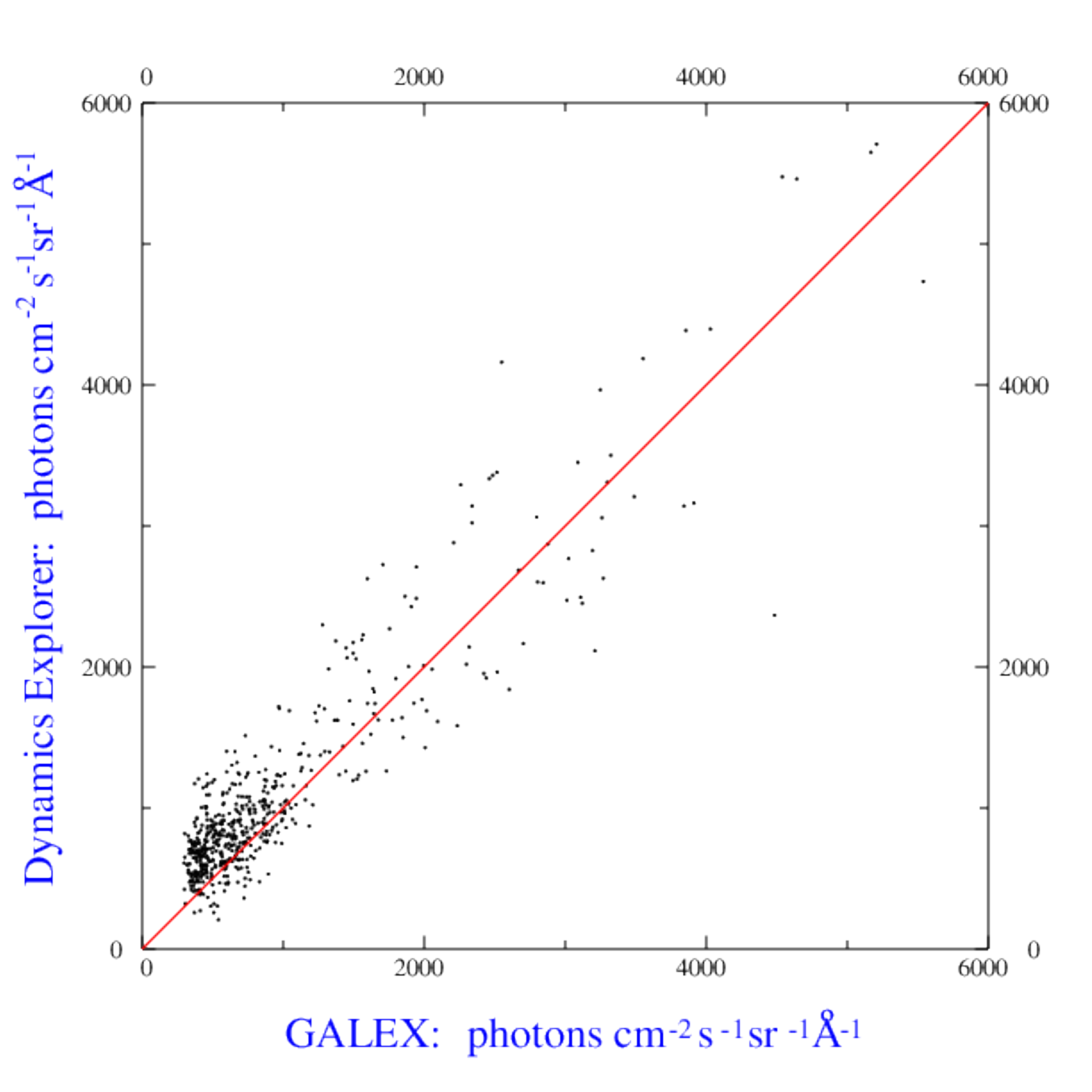}
\caption{That the GALEX observed Galactic Polar glow of $\sim 300$ photon units is unlikely to be of geophysical origin is supported by this quantitative comparison with the Dynamics Explorer FUV scans (that were shown in Figure 4). Each point represents multiple (mostly 3 or 4) observations of a GALEX target by DE, with the observations averaged to improved the statistics. \label{fig5}}
\end{figure}
\clearpage

\begin{figure} 
\epsscale{1.0}
\plotone{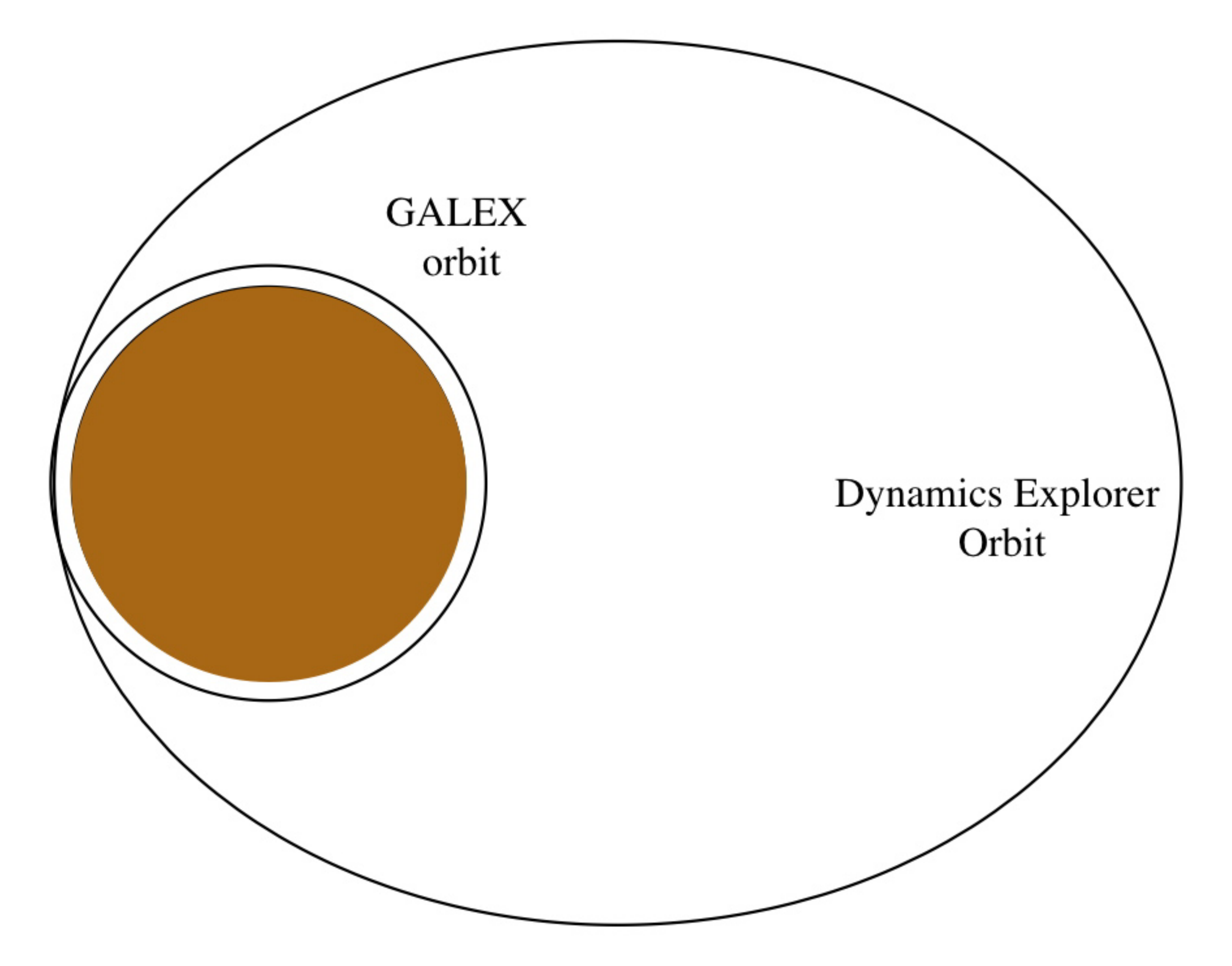}
\caption{ The GALEX orbit was nearly circular, but Dynamics Explorer was in an elliptical orbit having a distant (23,250 km) apogee where terrestrial UV emissions are expected to be very low indeed.  Thus, the DE observations of the diffuse ultraviolet background are of crucial importance in helping establish the astrophysical nature of most of the high-galactic-latitude diffuse FUV backgrounds that were observed by GALEX.  \label{fig6}}
\end{figure}
\clearpage

\begin{figure} 
\epsscale{0.9}
\plotone{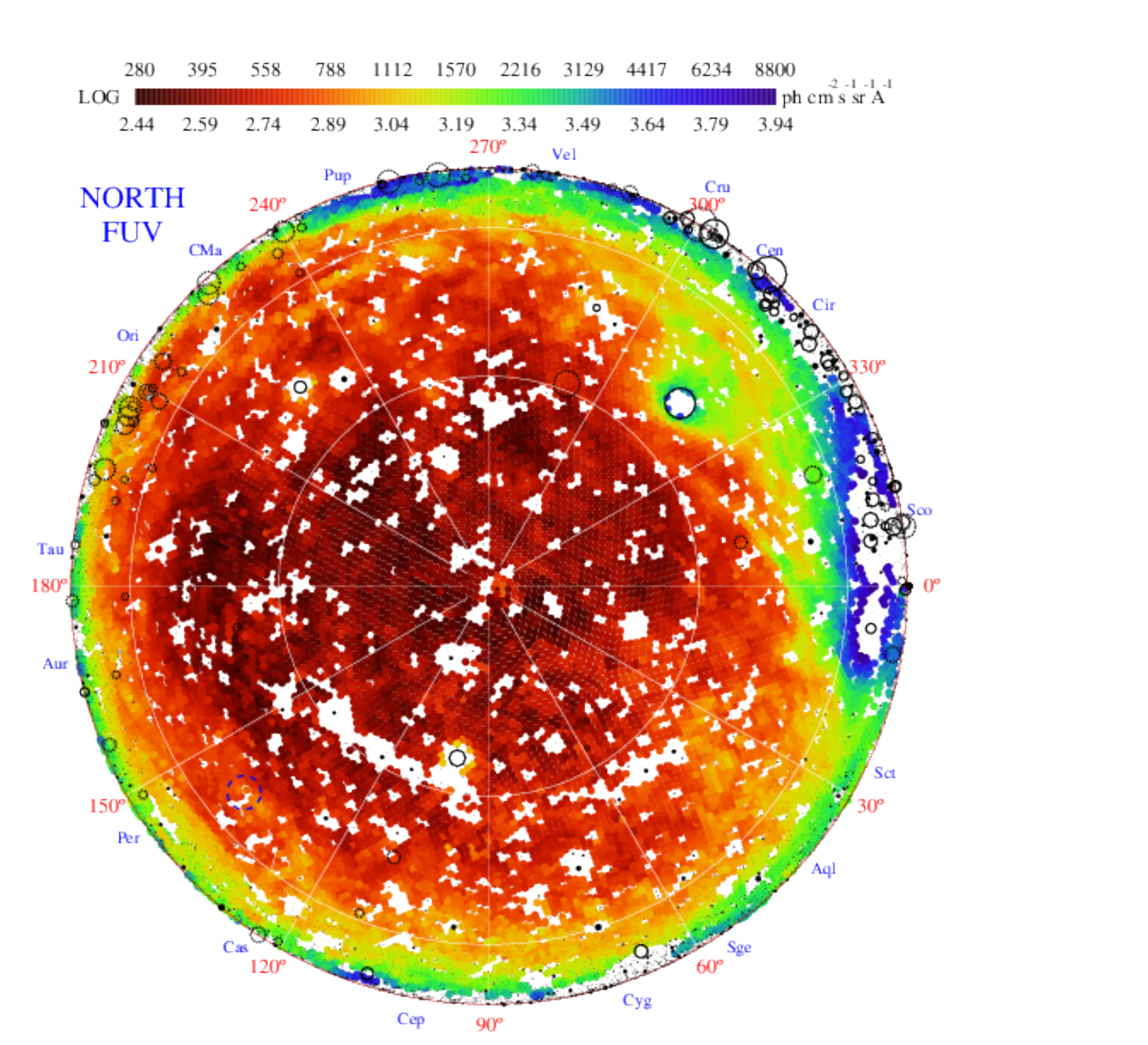}
\caption{GALEX FUV cosmic background brightness, on a logarithmic scale, for the Northern Galactic hemisphere.  (Regions where no GALEX images are available are white.)  Circles are UV-bright stars ({\em dashed} circles are the brightest UV stars in the {\em Southern} Galactic hemisphere; see Figure 11); they are highly concentrated around the upper rim of the figures.  The dust-scattered UV light of Spica (Murthy \& Henry 2011) is apparent at the upper right.  The image is dominated by a very-low-brightness UV glow (red) that shows almost no dependence on galactic longitude. The dimmest regions (black) are 280 photon units. This image alone can dispose of the notion that the FUV background at high galactic latitudes, if astrophysical, is due to dust-scattered starlight:  its origin is therefore a profound mystery.  The dashed circle at $\ell = 140^\circ, b=+40^\circ$ indicates the location of the reflection nebula discovered by Sandage (1976).  \label{fig7}}
\end{figure}
\clearpage

\begin{figure} 
\epsscale{1.10}
\plotone{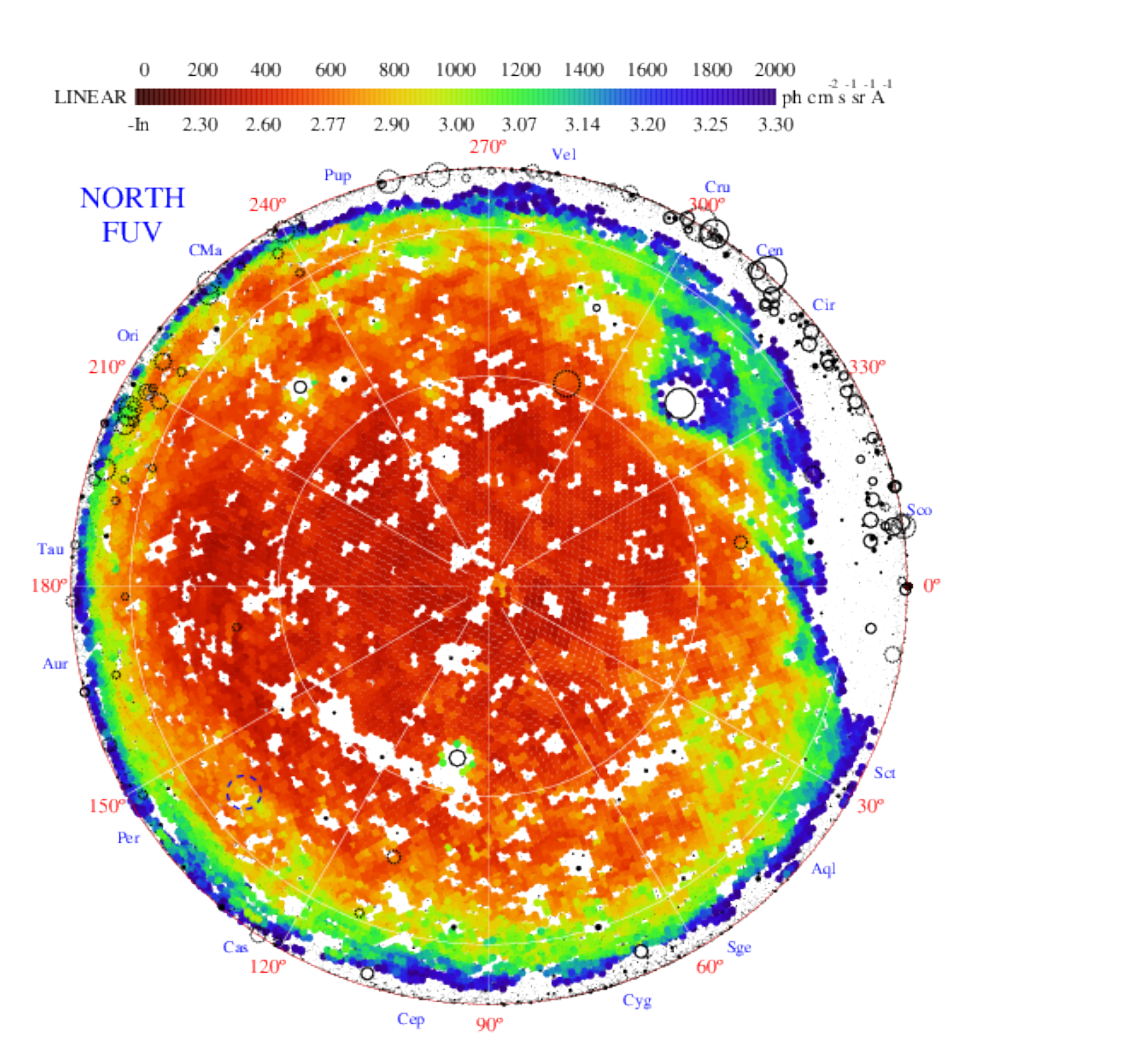}
\caption{ The same GALEX FUV brightness as in the previous figure, but this time on a {\em linear} scale, from zero, for the Northern Galactic hemisphere.  (Low-galactic-latitude regions that are brighter than 2000 units in the FUV are white.)    The dust-scattered UV light of Spica (Murthy \& Henry 2011) is again apparent.  One virtue of the linear scale is that the brightness and the relative uniformity of the galactic-cap FUV glow are more clearly seen, and are striking.  \label{fig8}}
\end{figure}
\clearpage

\begin{figure} 
\epsscale{1.10}
\plotone{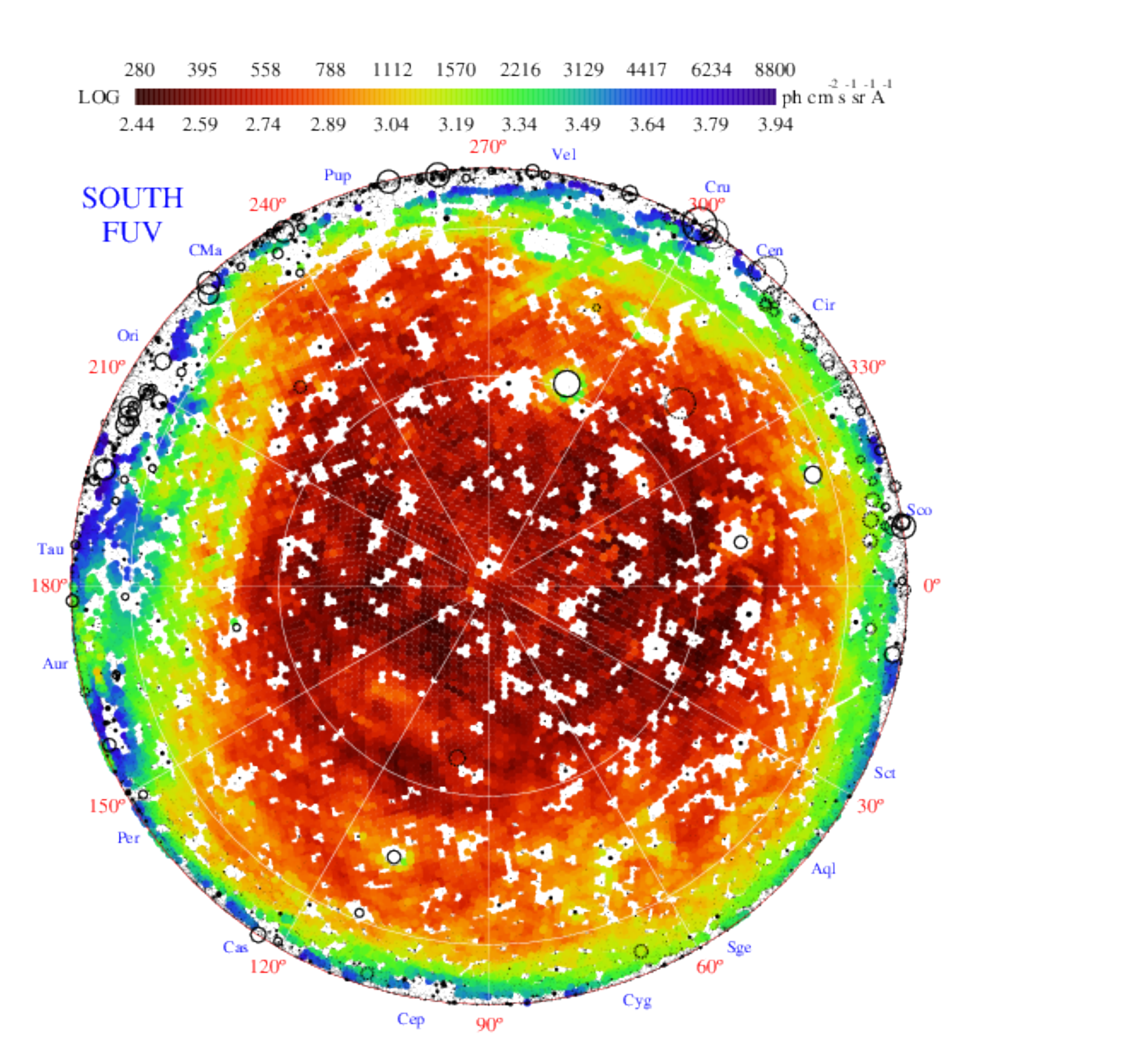}
\caption{  The same logarithmic scale as used in Figure 7, but this time for the {\em Southern} Galactic hemisphere.  The UV stars are here shown again , but this time it is the brightest ones (only) in the {\em Northern} Galactic hemisphere that are shown as dashed circles. Note the feature south of $-60^{\circ}$ between longitudes $120^{\circ}$ and  $150^{\circ}$ ; it is likely dust-scattered starlight.   \label{fig9}}
\end{figure}
\clearpage

\begin{figure} 
\epsscale{1.10}
\plotone{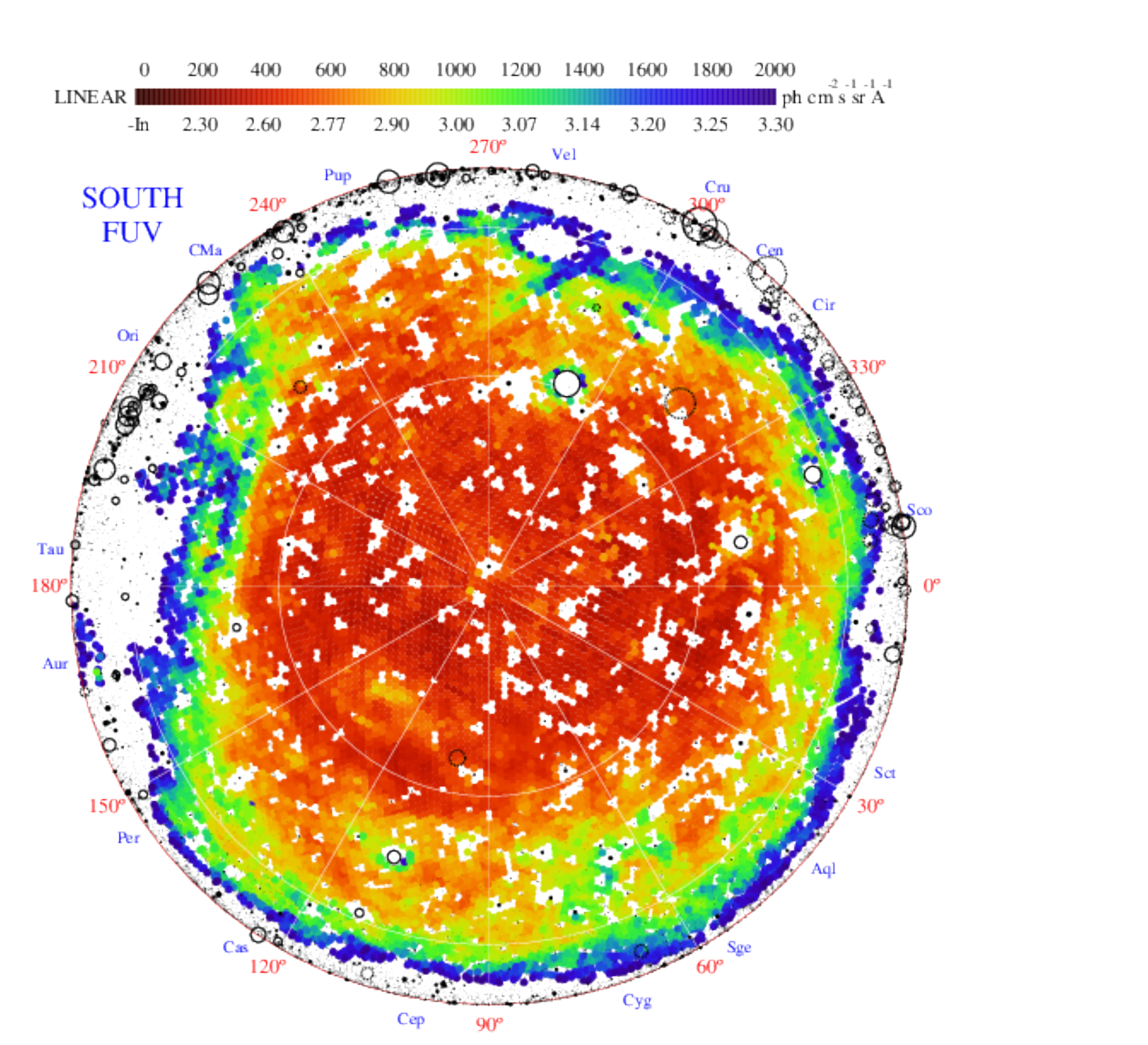}
\caption{ The same as Figure 9, but this time a {\em linear} plot for the Southern Galactic Hemisphere.  Again the remarkable uniformity of the general glow, with little or no Galactic longitude dependence, is striking.  Note, in this figure and in the previous one, the structure that appears between $120^\circ$ and $140^\circ$ longitude, south of the south $60^\circ$ latitude line:  this structure will also be seen in the infrared images, in following figures, suggesting, for these features, an origin in dust-scattered starlight.   \label{fig10}}
\end{figure}
\clearpage

\begin{figure} 
\epsscale{1.10}
\plotone{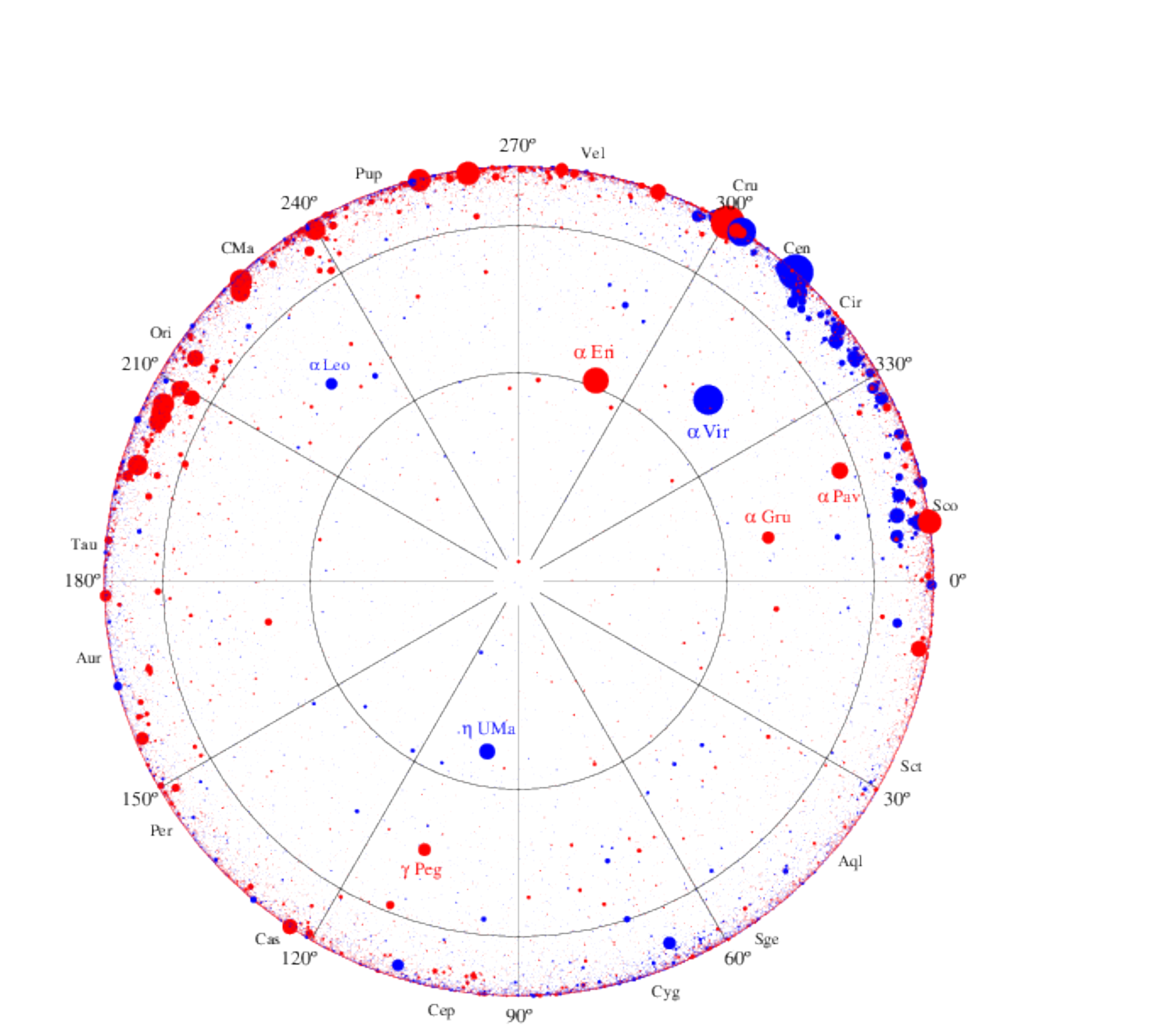}
\caption{ In previous figures, the locations of ultraviolet-bright stars were shown.  The same are shown again here, for both Northern (blue) and Southern (red) Galactic Latitudes.  The sky presents a strikingly different appearance in the far ultraviolet (compared with the visible) with almost all bright stars strongly concentrated not only toward the Galactic plane, but also toward that half of the Galactic plane lying between longitudes $180^\circ$ and $360^\circ$ (Henry 1977). \label{fig11}}
\end{figure}
\clearpage

\begin{figure} 
\epsscale{1.10}
\plotone{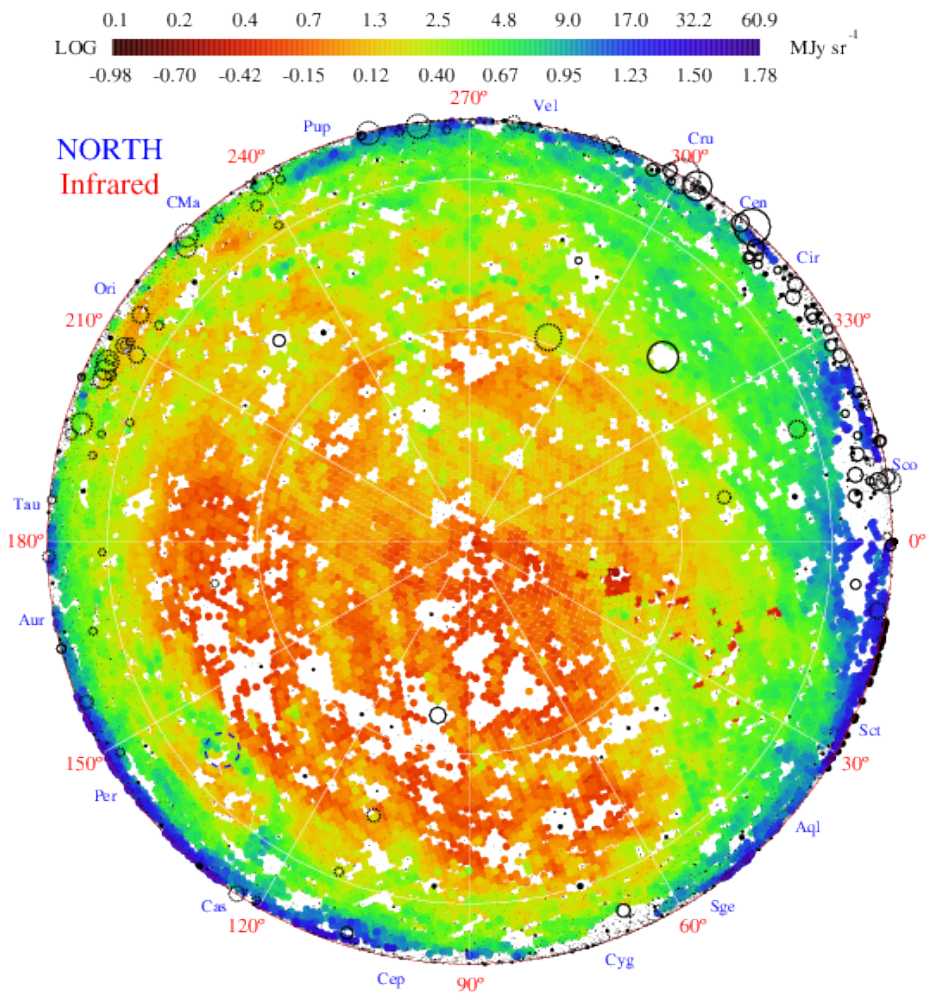}
\caption{The Northern Galactic hemisphere infrared diffuse background at $100\ \mu$m, shown here, using a logarithmic intensity scale, provides crucial assistance in interpeting the observed ultraviolet background.  This infrared background is widely thought to be (and we agree) due to simply thermal emission from interstellar dust that has been heated by starlight.  The plotted observations are from Schlegel et al. (1998).  Unlike the ultraviolet background (which was displayed in previous figures), in the infrared there is a strong {\em asymmetry} with galactic longitude, and hence a strong asymmetry in the interstellar dust distribution:  the {\em least} dust, is present in the longitude range $30^\circ - 210^\circ$, in the {\em northern} galactic hemisphere.   However, a quite {\em different} asymmetry (but an asymmetry that is just as strong) appears in the {\em southern} Galactic hemisphere$100\ \mu$m observations (Fig. 13). \label{fig12}}
\end{figure}
\clearpage

\begin{figure} 
\epsscale{1.10}
\plotone{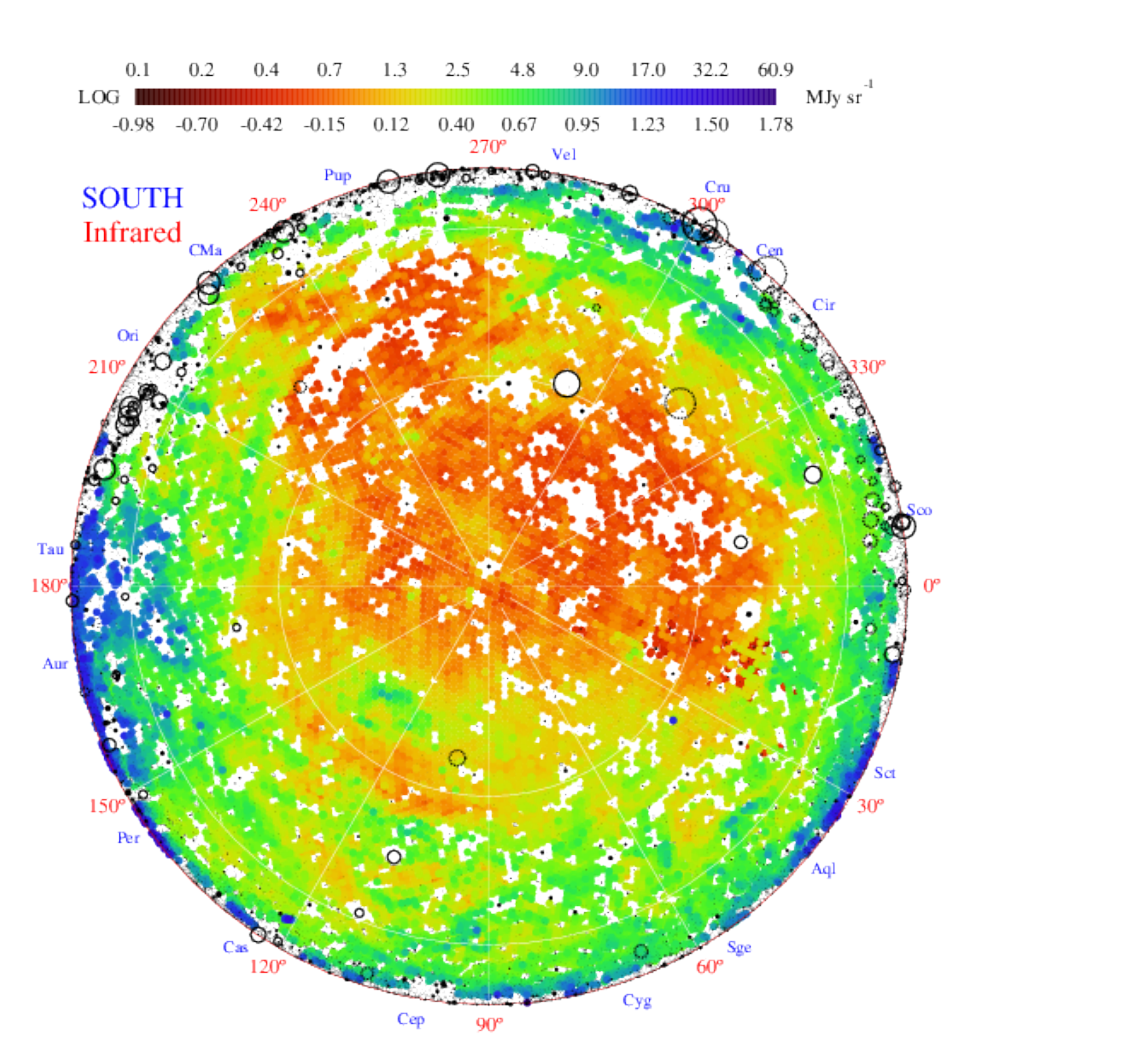}
\caption{ The Southern Galactic hemisphere $100\ \mu$m emission is shown using a logarithmic intensity scale.  Here, it is the (very different) longitude range $210^\circ - 0^\circ - 30^\circ$ that has the least amount of interstellar dust.  This is in (drastic) contrast with the {\em independence} of Galactic longitude, of the far-ultraviolet background radiation that was demonstrated in Figures 7-10. \label{fig13}}
\end{figure}
\clearpage

\begin{figure} 
\epsscale{1.0}
\plotone{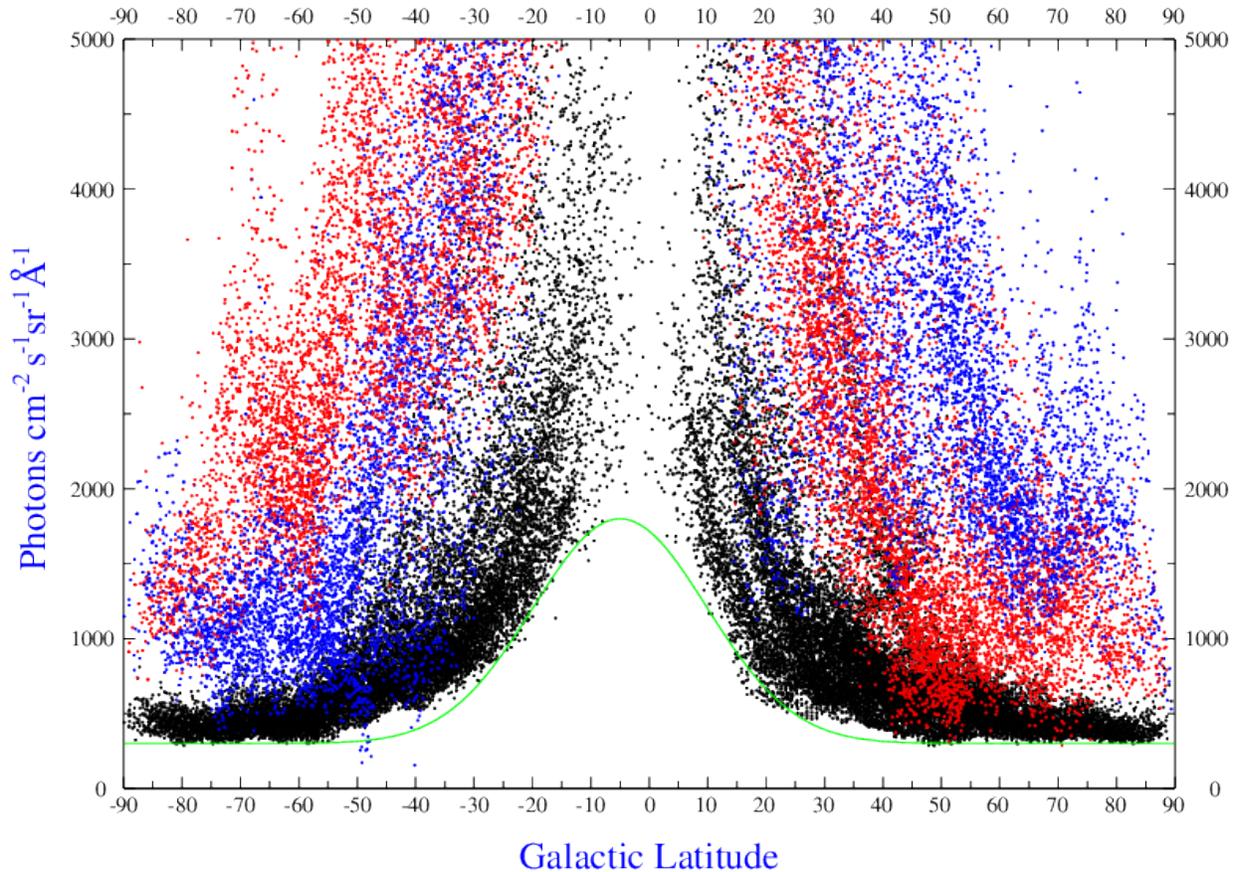}
\caption{The Galactic Latitude dependence of the far-ultraviolet background (black), and of the $100\ \mu$m emission (red, for the Galactic Longitude range $30 ^\circ - 210 ^\circ$; blue, for the Galactic Longitude range $210 ^\circ - 0 ^{\circ} - 30 ^\circ$).  There is a dramatic difference in the latitude dependences of the $100\ \mu$m emission  for these two longitude ranges, which is {\em not} true for the FUV diffuse emission (see Figure 15), which therefore must have an independent origin.  (The green curve (symmetric about $-5^\circ$ Galactic Latitude)  approximates the lower bound of the FUV emission.)  This figure establishes that there is no connection between the diffuse infrared emission and most of the diffuse FUV emission; the source of the diffuse FUV emission is unknown---that is the ``Mystery" that is referred to in the title of this paper.   \label{fig14}}
\end{figure}
\clearpage

\begin{figure} 
\epsscale{1.0}
\plotone{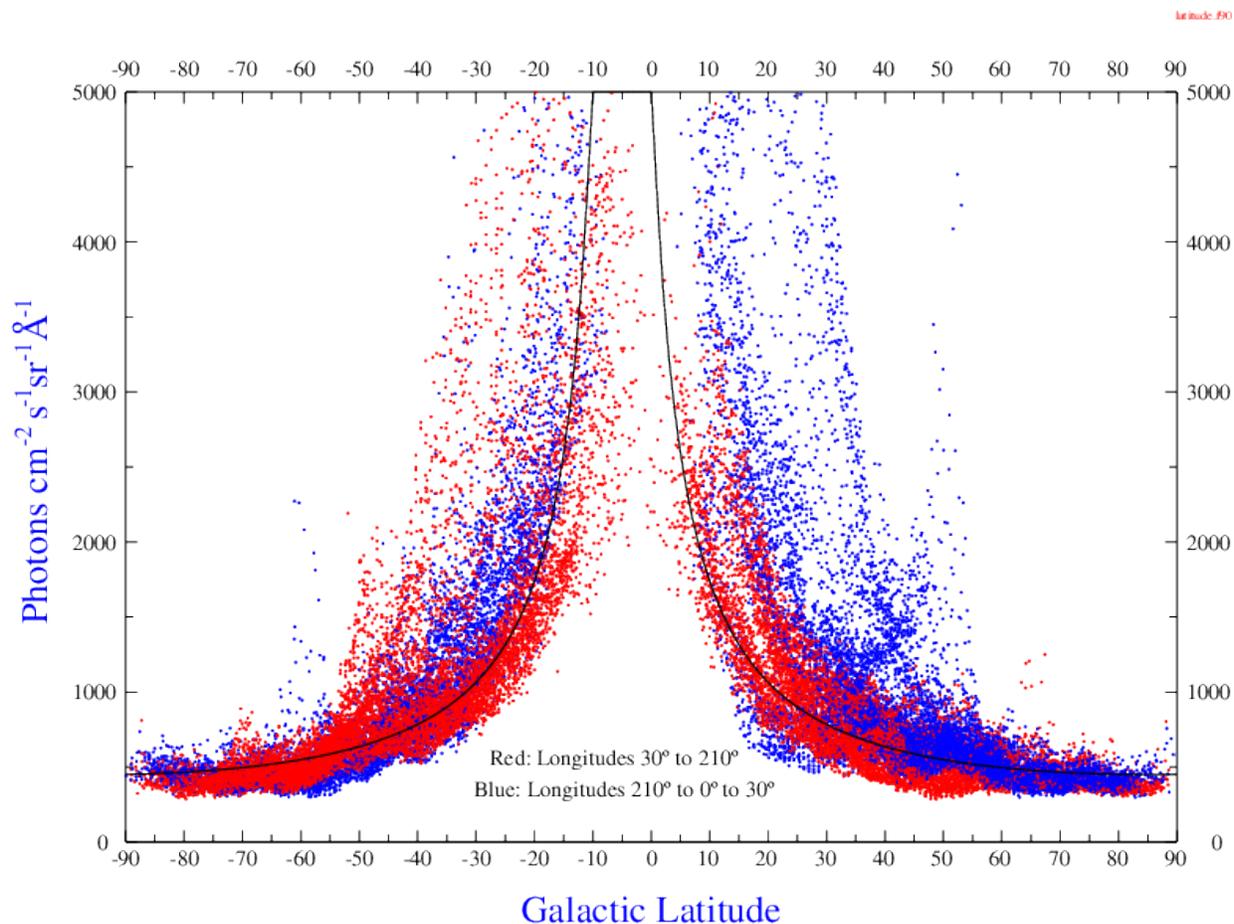}
\caption{The solid black line is $csc(b-5^\circ)$, where {\em b} is the Galactic Latitude.  The FUV brightnesses have here been subdivided according to Galactic Longitude (exactly as was done for the $100\ \mu$m emission in Figure 14) in order to bring out the fact that there is little or no sign of the strong Galactic Longitude asymmetry that appears in the $100\ \mu$m emission, as displayed in Figure 14.  Also, the fact that the FUV brightness follows the black line strongly suggests that at least most of the FUV emission originates in a layer that is centered on the Galactic plane.  \label{fig15}}
\end{figure}
\clearpage

\begin{figure} 
\epsscale{0.8}
\plotone{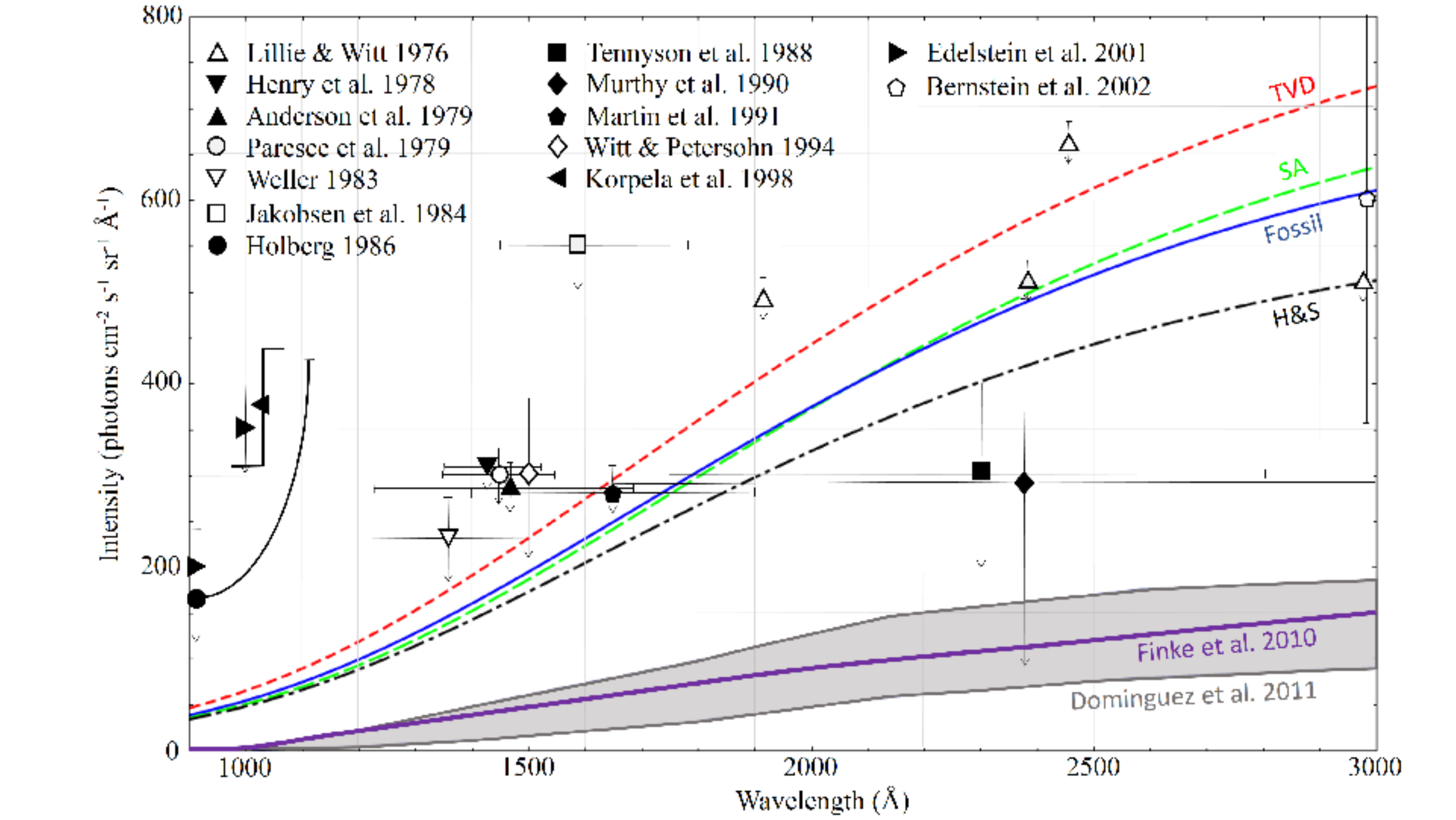}
\caption{Curves give model predictions for the expected diffuse ultraviolet background, in photon units, on the assumption that the highest Galactic latitude background is entirely due to the integrated light of distant galaxies or extragalactic background light (EBL).  Shown are predictions by Finke et al. (2010, ``Model C") and Dominguez et al. (2011, upper and lower limits).  Also shown are upper limits obtained by adapting a code for EBL intensity at near-optical wavelengths (Overduin, Prins, \& Strobach 2014). 
To generate the latter curves, we have used templates for both quiescent and star-forming galaxy spectra over the full spectrum from FUV to sub-mm wavelengths (Devriendt et al. 1999).
Evolution in galaxy luminosity and number density is incorporated by requiring that the overall luminosity density of the universe be consistent with theoretical and observational constraints at every redshift, as compiled by Nagamine et al. (2006).
The labels TVD, SA, Fossil and H\&S refer to specific evolution models, and parameters within each model have been adjusted to give the largest possible spread in predicted EBL intensities.
We have incorporated a model for extinction by dust in the intergalactic medium due to Loeb \& Haiman (1997).  All curves assume a  standard $\Lambda$CDM cosmology (with $\Omega_{M}=0.3$ and $\Omega_{\Lambda}=0.7$).
Superimposed on these predicted EBL intensity curves are forty years' worth of observational constraints (datapoints and error bars), spectroscopic (filled symbols) as well as mostly photometric (empty symbols).
The GALEX FUV data on which we focus in this paper gives a minimum background of about 300 photon units over the range 1350 \AA\ - 1750 \AA.
\label{fig16}}
\end{figure}
\clearpage

\begin{figure} 
\epsscale{1.0}
     \plotone{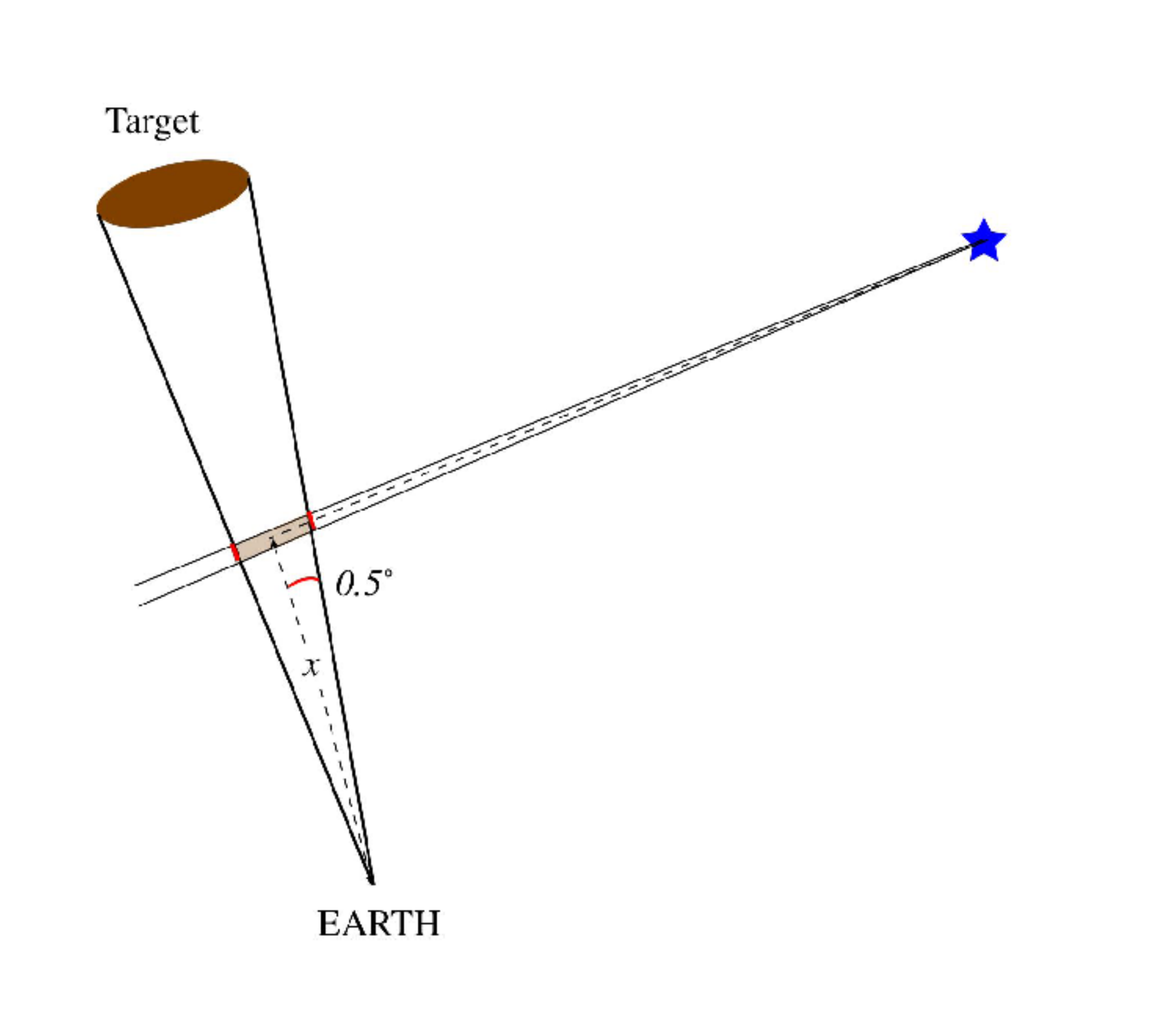}
\caption{The geometry for our simple single-scattering model giving the expected flux from the dust-scattered light of stars.  The GALEX field of view is one degree.  For each star contributing dust-scattered light to our observation of the given target, we make three calculations:  1) the diminution of intensity of the light of each star (one typical star is shown) due to both its distance from each location on our line of sight, and the attenuation due to absorption by dust along the path to our observing line of sight,  2) the amount of light scattered in our direction as the starlight crosses the slice of our line of sight, and  3) the diminution again, as in the first step, but now from the scattering location to our observing location at Earth.   \label{fig17}}
\end{figure}
\clearpage

\begin{figure} 
\epsscale{1.0}
     \plotone{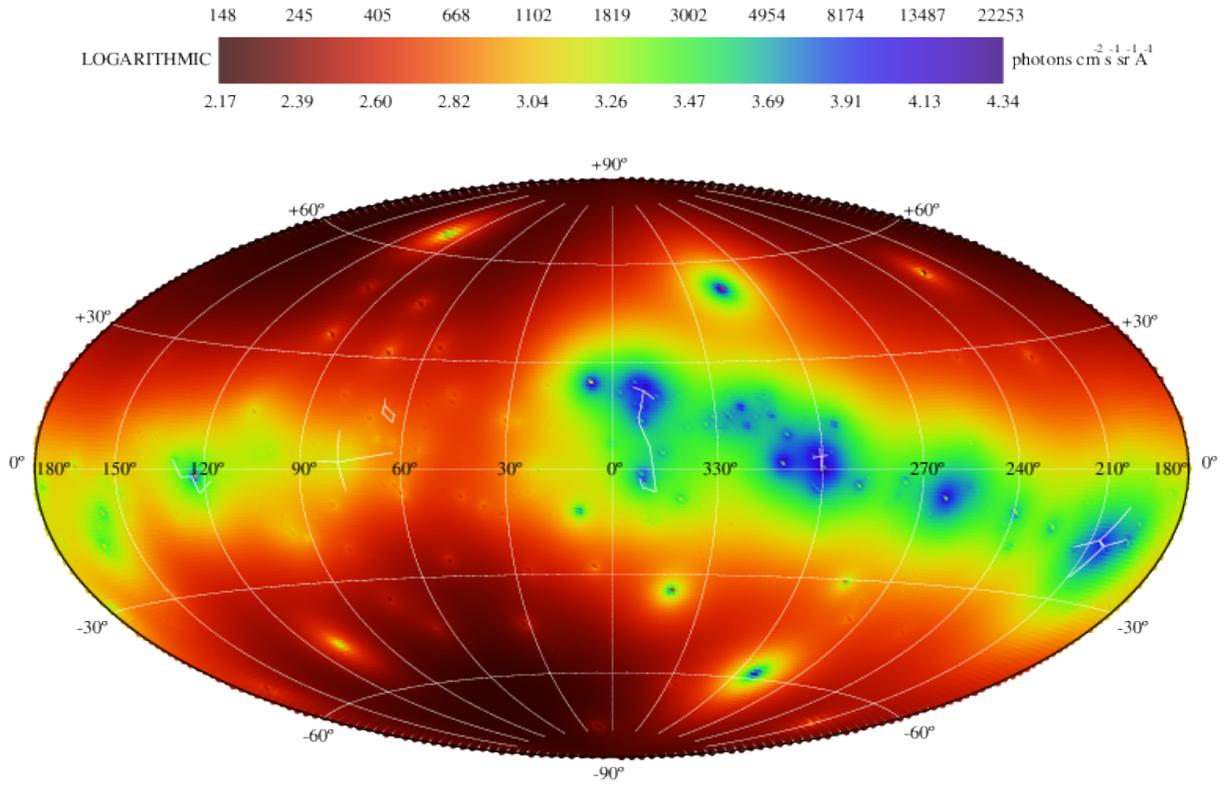}
\caption{Our model for the 1550 \AA\  diffuse background, predicted using the values of the albedo and scattering parameter {\em g} found by Hamden, Schiminovich, \& Siebert (2013).   A few constellation patterns are included for orientation. The brightest spot predicted is at $\ell = 299.8, b = -1.0$, which is in the Coalsack nebula in the constellation Crux.  This figure shows our prediction for the entire sky, whereas the following figure is the same model, but limited to the regions that were observed in the FUV by GALEX.  \label{fig18}}
\end{figure}
\clearpage

\begin{figure} 
\epsscale{1.0}
     \plotone{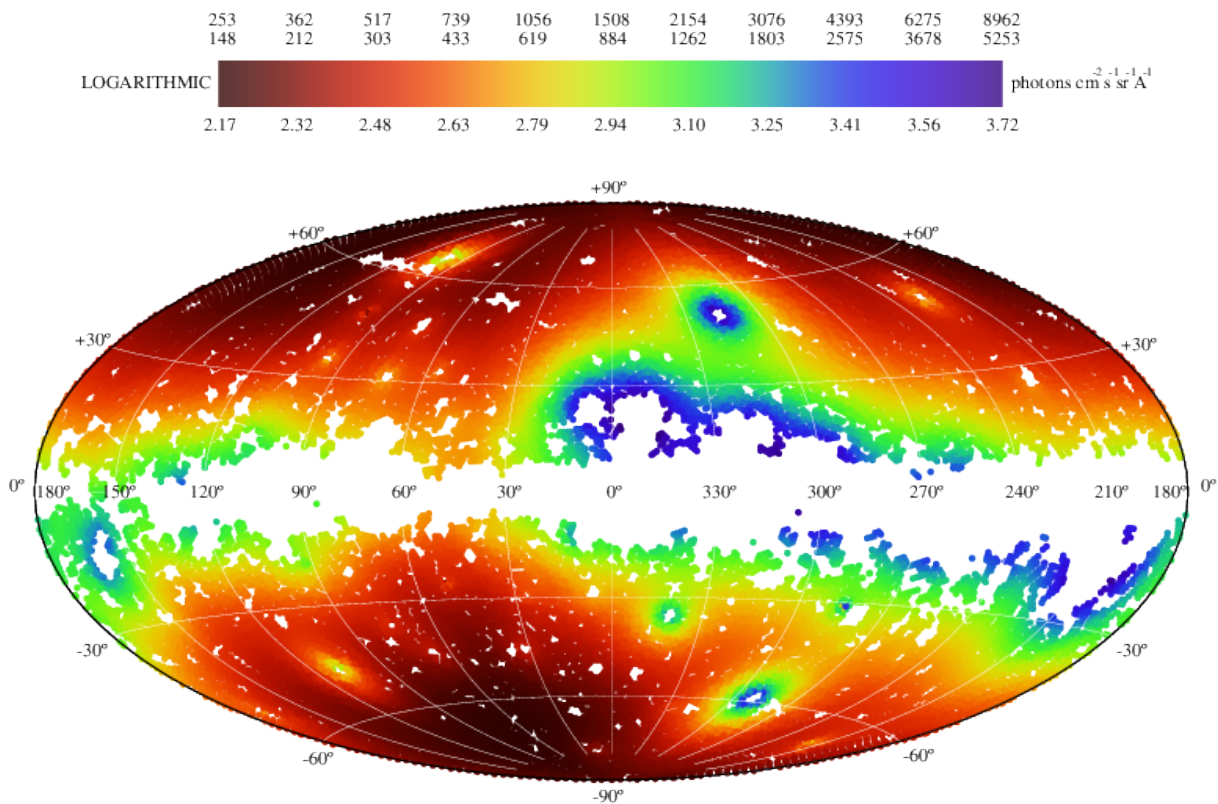}
\caption{Our prediction of the diffuse far-ultraviolet background assuming that it is entirely due to the light of FUV-bright stars scattered by interstellar dust having the albedo (0.62) and the Henyey-Greenstein scattering parameter (0.78) that were found by Hamden et al. (2013) to fit the GALEX data.   The upper scale at the top of the figure is equal to the actual model scale (given immediately below it), simply multiplied by 1.706 to force agreement with the data (Figure 17) at the brightest spot. Note the extraordinarily low predicted FUV brightness between Galactic longitudes  30$^\circ$ and  90$^\circ$, at the lowest Galactic latitudes.  In Figure 20 we again present the data, for comparison with this model, and in the figure following we give an adjusted plot of the ratio (observed/model).    \label{fig19}}
\end{figure}
\clearpage

\begin{figure} 
\epsscale{1.0}
     \plotone{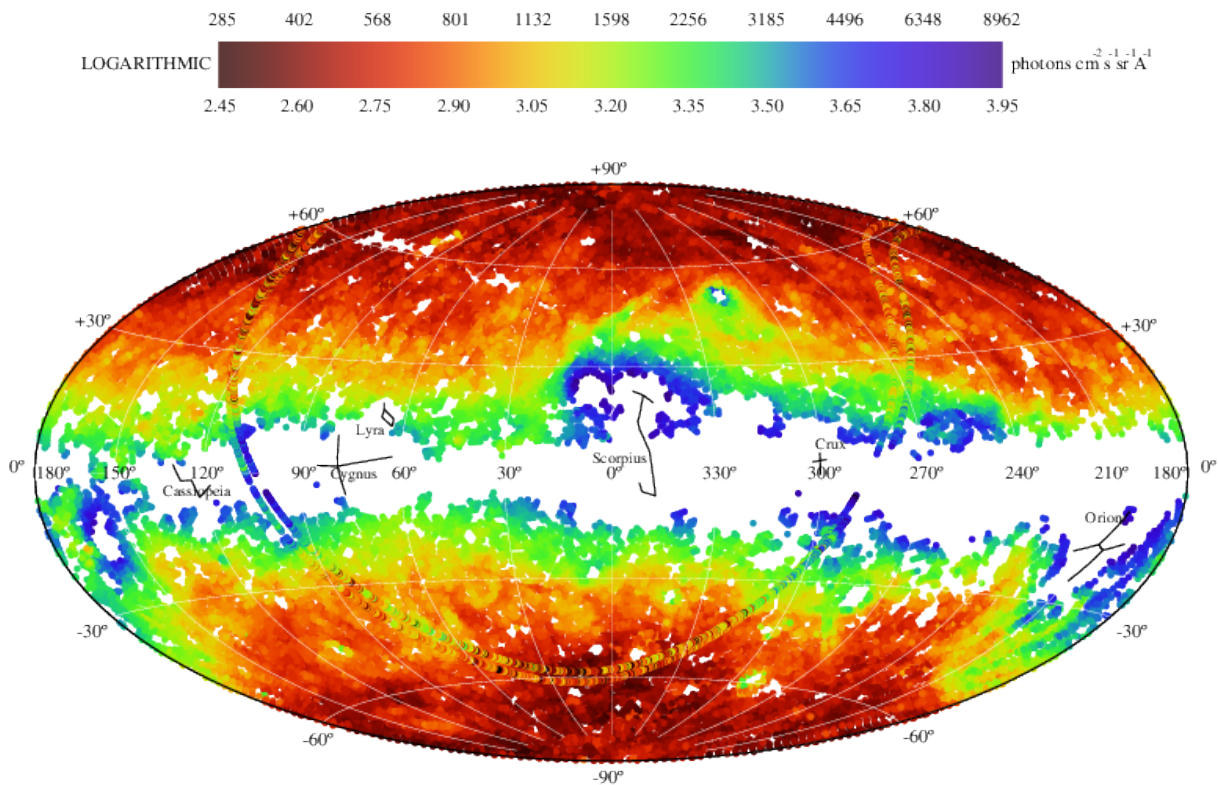}  
\caption{This figure is the same as Figure 1, but with the ``supposed-source" star locations no longer marked with blue dots.  This permits us to see whether individual isolated bright stars do give rise to a diffuse background.  The relevant stars are listed in Table 2.  We see that only in the case of Spica is there an extended scattered-light contribution to the observed diffuse FUV background.  For Achernar, there is scattered FUV, but only very close to the star's location.  In all other cases, it is difficult or impossible to attribute the diffuse background to a particular source star.  Stars that are detected have been analyzed by Murthy \& Henry (2011). (This figure also includes the Dynamics Explorer scans, on exactly the same intensity scale.  Note, particularily, the two Galactic-plane crossings between Cassiopeia and Cygnus, showing strong diffuse FUV emission at the lowest Galactic latitudes.)  \label{fig20}}
\end{figure}
\clearpage

\begin{figure} 
\epsscale{1.0}
     \plotone{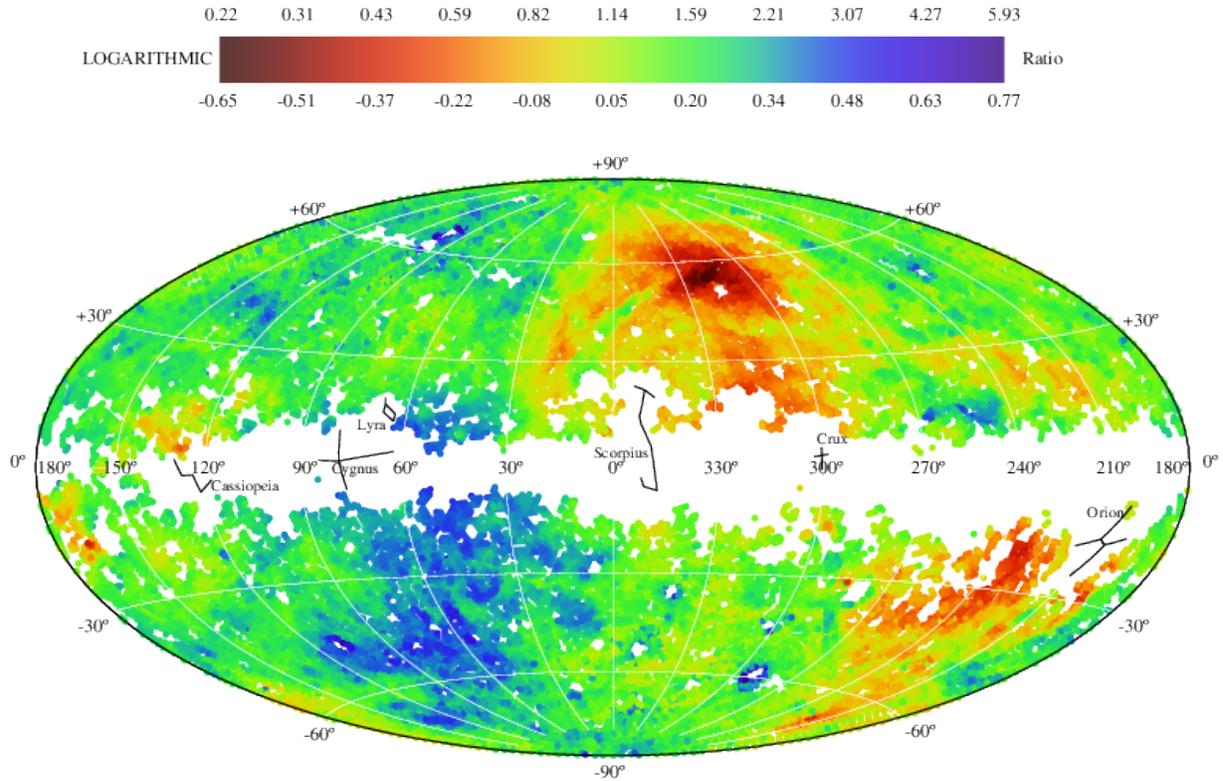}   
\caption{How well does our dust-scattered FUV model fit the Murthy, Henry, \& Sujatha (2010) observations?  This figure shows that (with important caveats) the fit is remarkably good!  What is plotted is the Ratio:  (FUV Observed)/(FUV Model)$\times 0.586$, where $0.586=5253/8962$ to normalize to highest brightness, and where in this case we have eliminated from our run of the model six stars (all of those in Table 2 except for Spica) because (Figure 20) those stars are seen to contribute little or nothing to the diffuse radiation that is observed.  Some of those omitted stars are detectable in this plot  as slight ``excesses" of observed/model, meaning of course that they are in fact detected (if only slightly) by GALEX.    \label{fig21}}
\end{figure}
\clearpage 

\begin{figure} 
\epsscale{1.0}
     \plotone{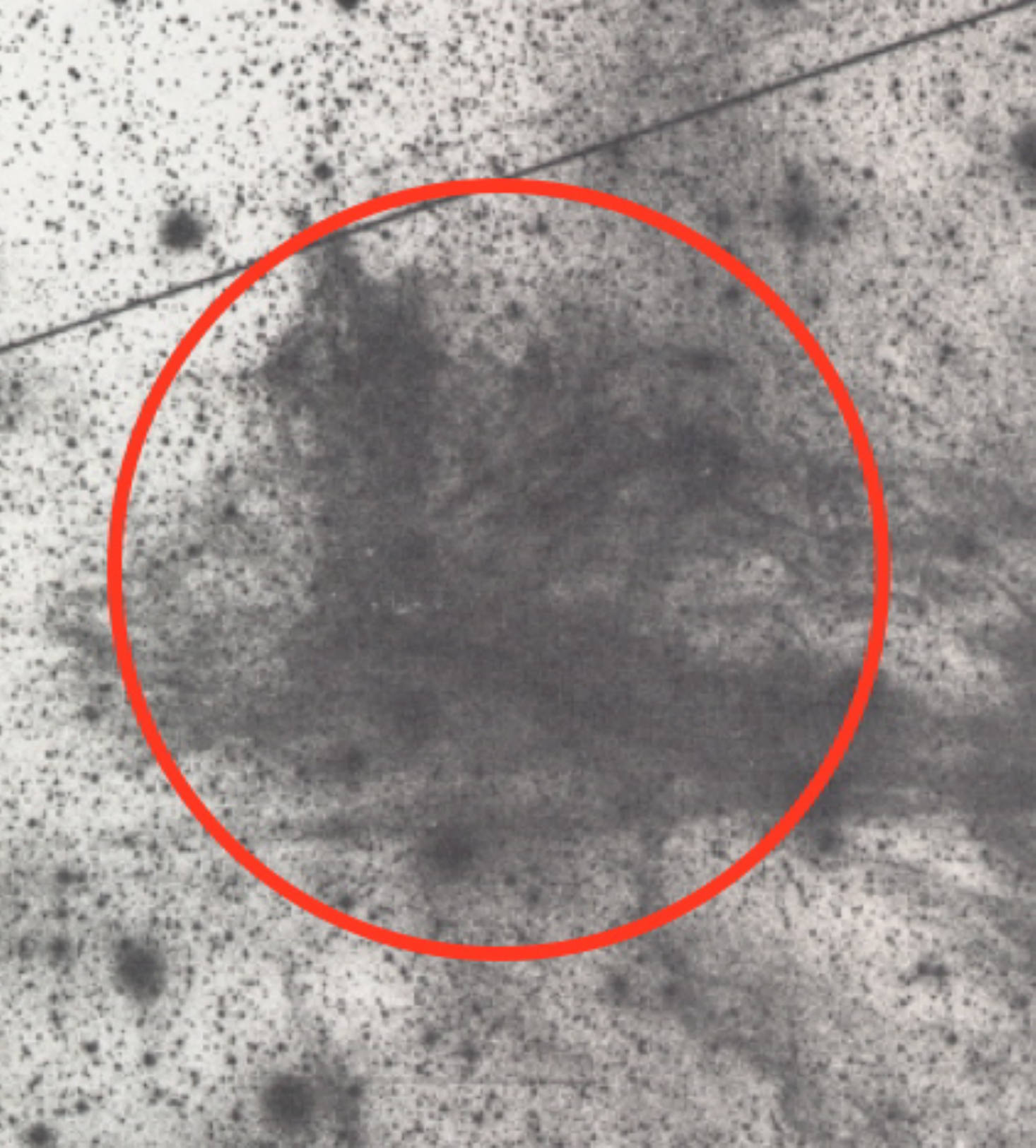}                
\caption{This is the visible-light image of the high-Galactic-latitude interstellar dust cloud that was discovered by Sandage (1976).  Note the airplane lights!  Henry and Murthy obtained GALEX images of part of this nebula (red circle) in their GALEX Guest Investigator program,  to test their hypothesis (which turned out to be incorrect) that the GALEX FUV radiation was extragalactic in its origin.  Detailed modeling of their observation is given by Henry (2010).  \label{fig22}}
\end{figure}
\clearpage

\begin{figure} 
\epsscale{1.0}
     \plotone{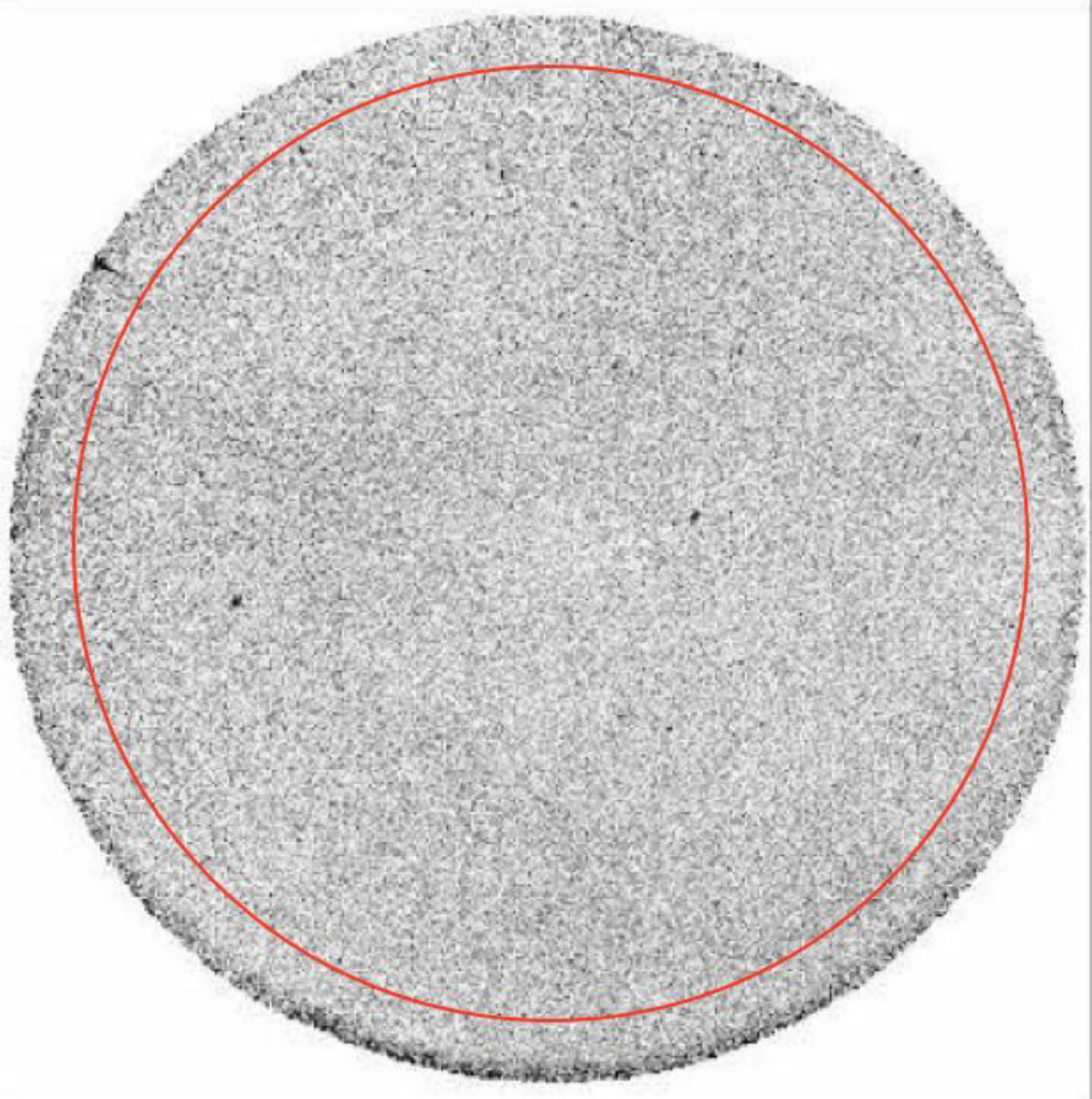}                   
\caption{This is the GALEX FUV image of the Sandage nebula for which a visible-light image is given in Figure 22. There is no resemblance at all between the visible and FUV images.  Only 6.6\% of the counts in this FUV image are due to stars, the rest is diffuse FUV radiation.  This picture is an image of our ``second component" of the FUV diffuse background radiation; its origin, is the ``mystery" referenced in the title of our paper.  The same image is presented in three dimensions in Henry (2010).   \label{fig23}}
\end{figure}
\clearpage

\begin{figure}[h!]
\begin{center}
\includegraphics[width=30mm]{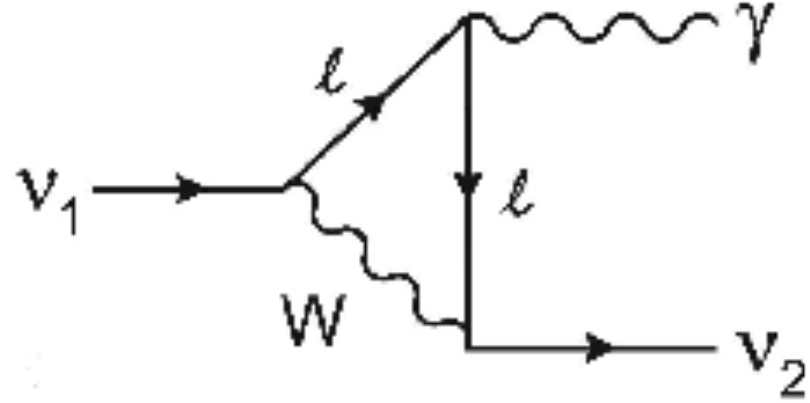}
\caption{Feynman diagram for decay of a massive neutrino into a lighter neutrino plus a photon.}
\label{fig-neutrino}
\end{center}
\end{figure}

\begin{figure}[h!]
\begin{center}
\includegraphics[width=70mm]{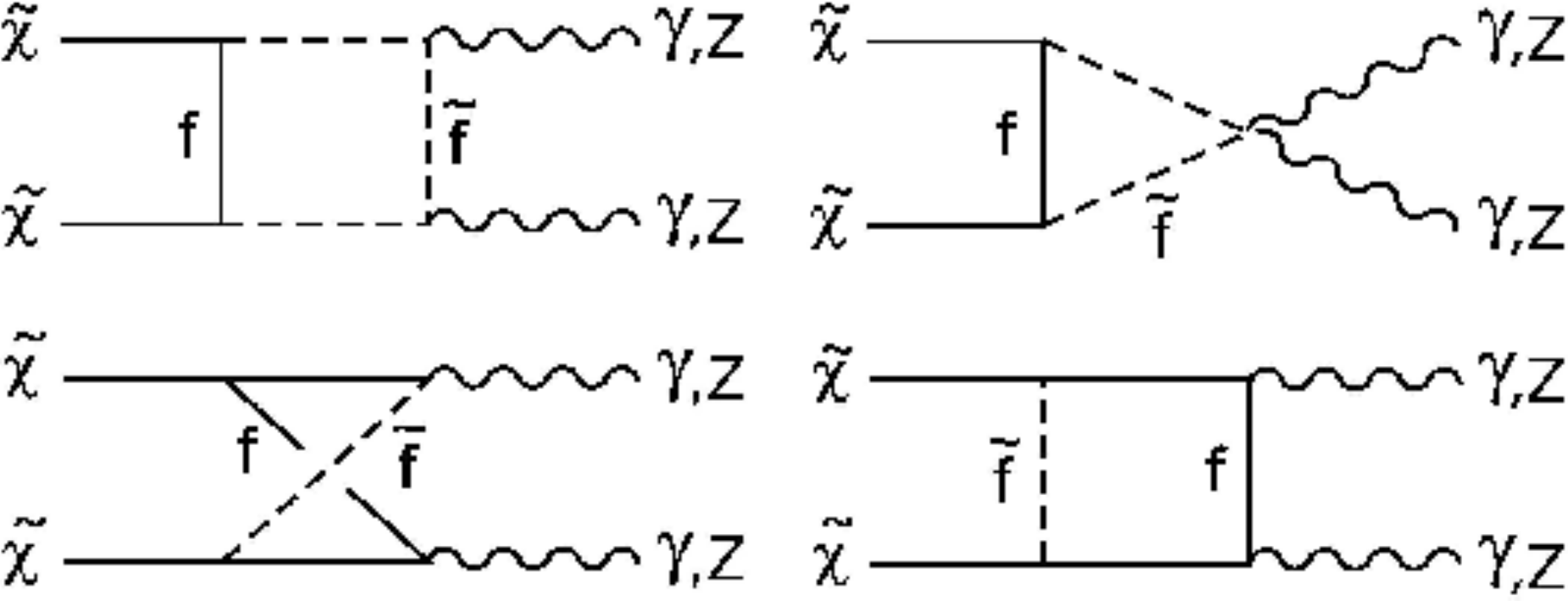}
\caption{Feynman diagrams for the annihilation of two WIMPs (here, supersymmetric neutralinos, $\tilde{\chi}$) into a photon pair (or a photon plus a Z boson) via intermediate fermions and their supersymmetric partners, the sfermions ($\tilde{f}$).}
\label{fig-wimpAnnihilation}
\end{center}
\end{figure}

\begin{figure}[h!]
\begin{center}
\includegraphics[width=35mm]{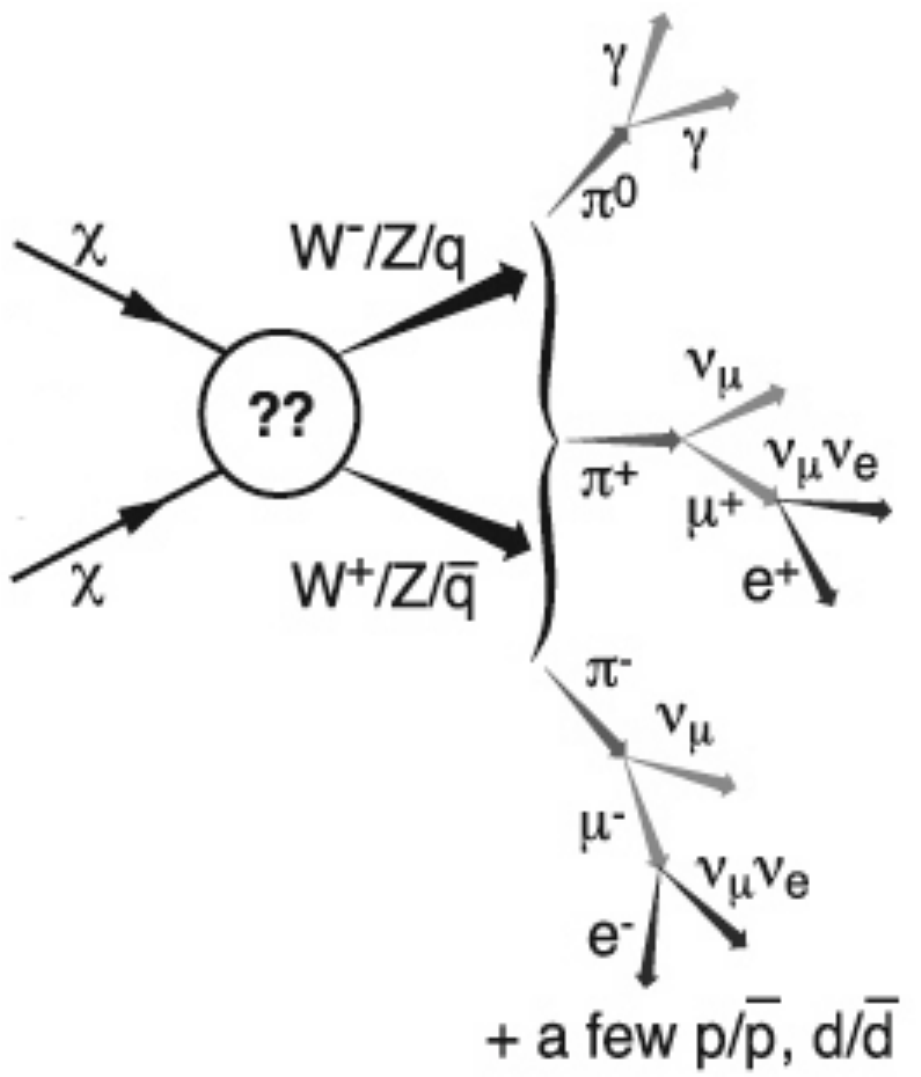}
\caption{Schematic depiction of tree-level WIMP annihilations to quarks and bosons, leading to showers of secondary photons, neutrinos and antimatter.}
\label{fig-wimpIndirect}
\end{center}
\end{figure}

\begin{figure}[h!]
\begin{center}
\includegraphics[width=85mm]{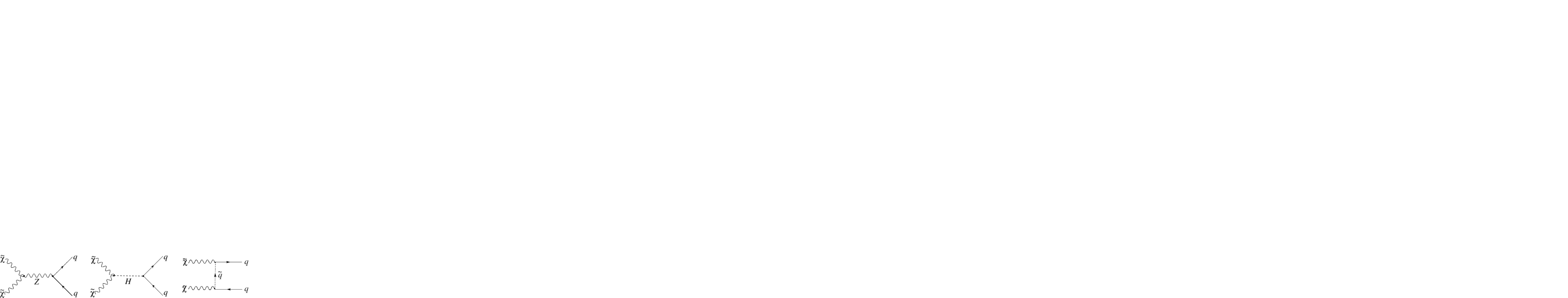}
\caption{Example Feynman diagrams for tree-level WIMP annihilations to quark-antiquark pairs via intermediate Z bosons, neutral Higgs bosons or squarks. These and similar processes occur inside the circle labeled ``??'' in Fig. 26.}
\label{fig-wimpTreeLevel}
\end{center}
\end{figure}

\begin{figure}[h!]
\begin{center}
\includegraphics[width=27mm]{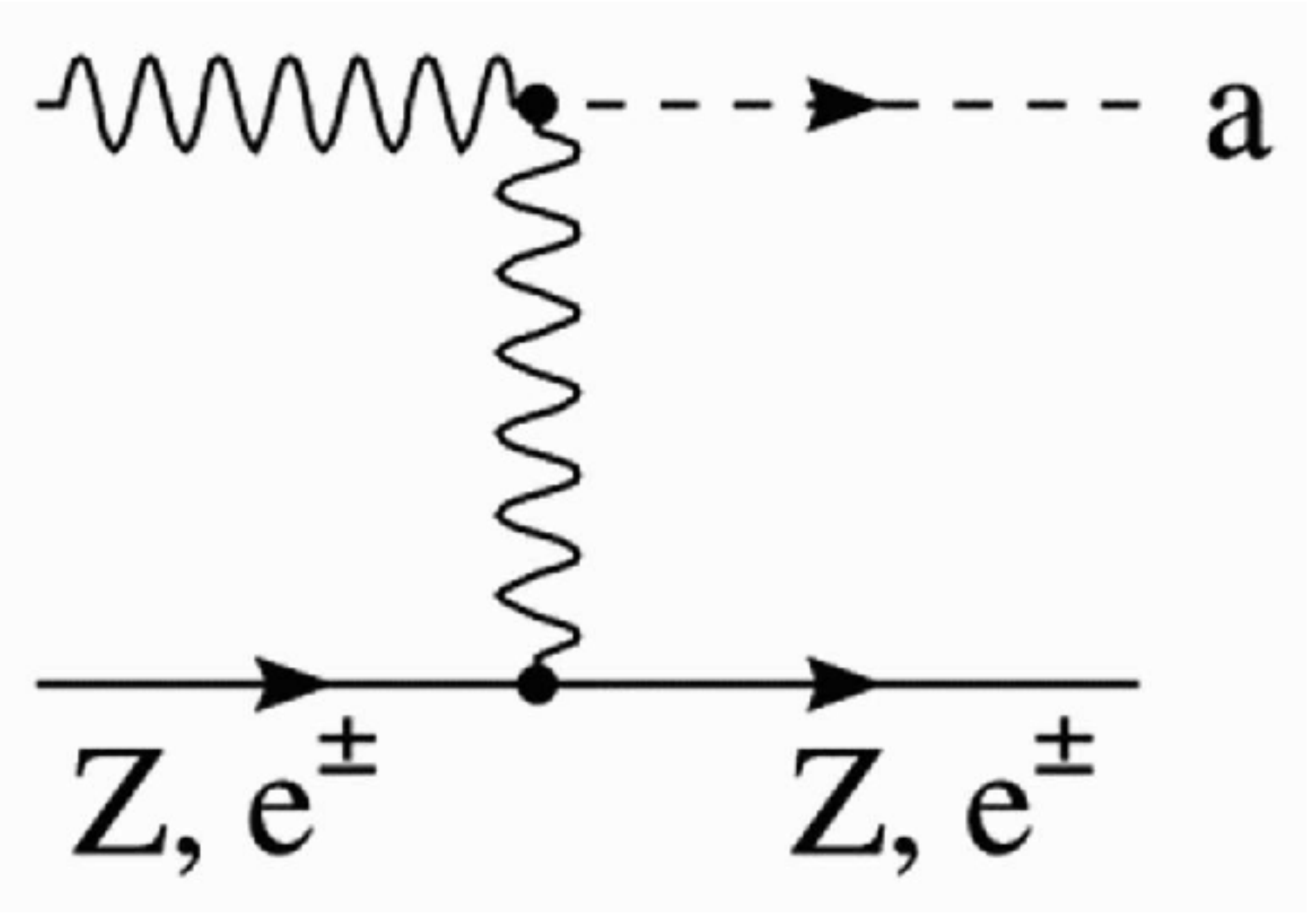}
\caption{Feynman diagram for photon-axion interconversion via the Primakoff effect.}
\label{fig-primakoff}
\end{center}
\end{figure}

\begin{figure}[h!]
\begin{center}
\includegraphics[width=27mm]{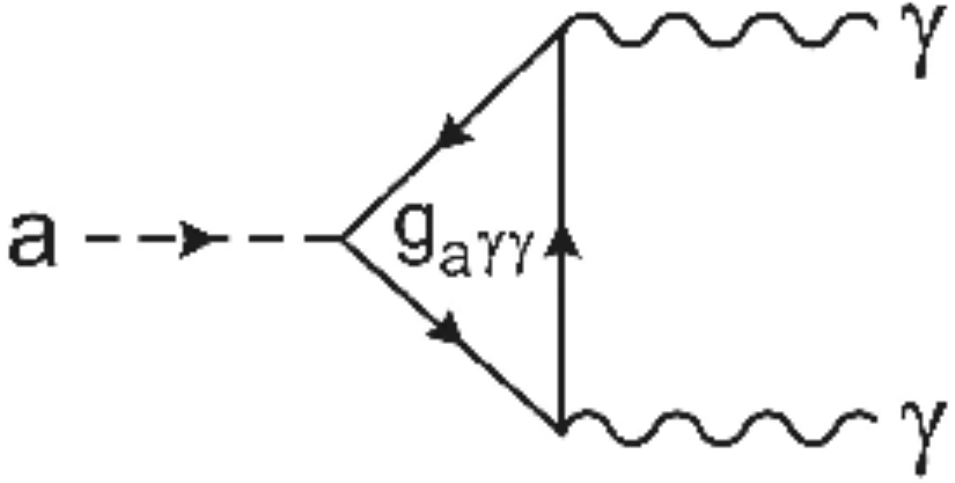}
\caption{Feynman diagram for the decay of light thermal axions to two photons via an intermediate fermion loop, characterized by a coupling constant $g_{a\gamma\gamma}$.}
\label{fig-axionDecay}
\end{center}
\end{figure}

\clearpage
\begin{figure} 
\epsscale{0.8}
     \plotone{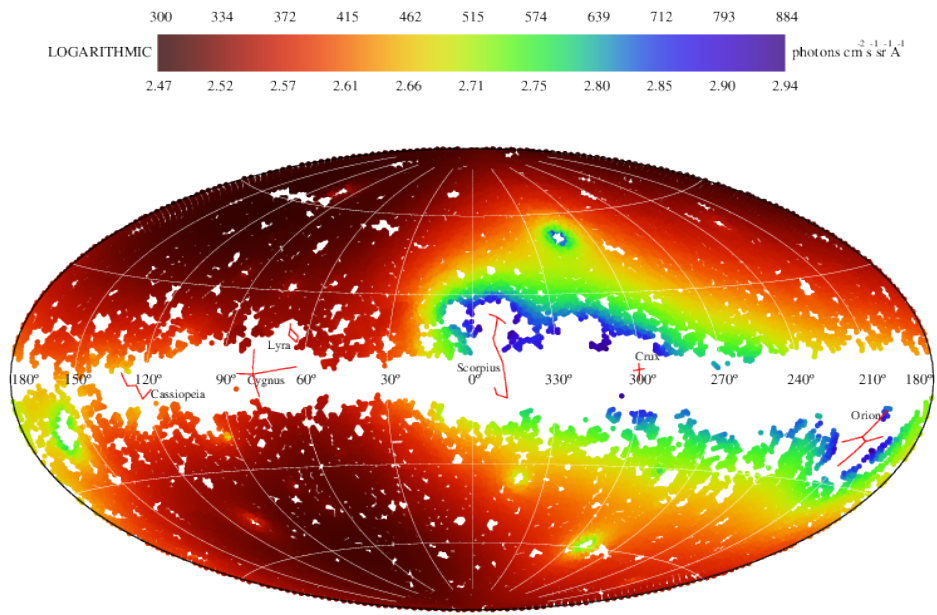}     
\caption{This is our model prediction for the case of interstellar grains that are much smaller than the wavelength of the radiation that is scattered.  Such grains will scatter the radiation isotropically and with unit albedo (Draine 2011).  Our model for forward-scattering grains (Figures 18 and 19), which is excellent except for its not taking into account local variations from place to place in the density of the scattering dust, has here been crudely adapted for the case of {\em isotropic} scattering.  We have forced the model to produce a high-Galactic-latitude brightness of 300 photon units (as observed) by simply reducing the density of dust in the interstellar medium by a factor of eleven.  That drastic reduction seems to be what is necessary to get the low observed background at high Galactic latitudes.  It probably simply accommodates the fact that we happen to be located in a very low-density region of the interstellar medium (Perry \& Johnston 1982).  An important point to note is that the ratio of the brightest predicted value to the lowest predicted value is a factor of only 2.95, whereas in Figure 1 the observed same ratio is a factor of 31.4!      \label{fig30}}
\end{figure}
\clearpage

\begin{figure} 
\epsscale{1.0}
     \plotone{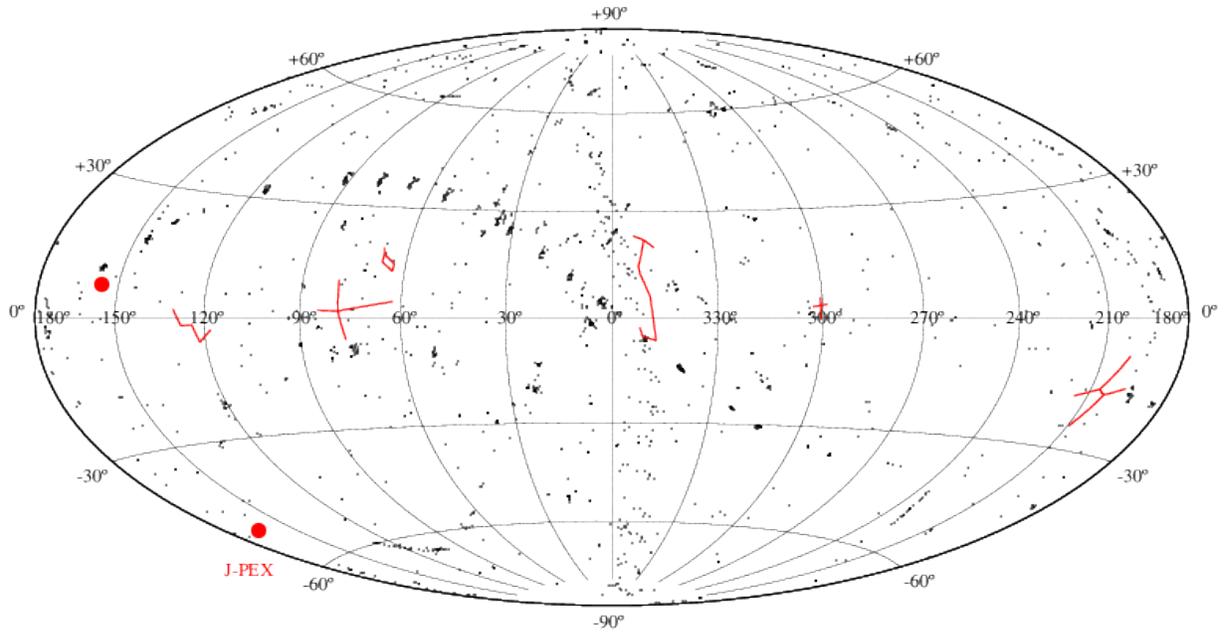}     
\caption{Locations (black dots) are shown for the 1943 {\em Voyager} observations of the diffuse ultraviolet background at wavelengths shorter than Lyman $\alpha$ that were reported by Murthy, Henry, \& Holberg (2012).  The red circle labelled J-PEX shows the location of the J-PEX observation (at 220 - 250 \AA) of Feige 24, reported by Kowalski et al. (2011), who found a mysterious bright diffuse background.  The other red circle is at the location of the J-PEX observation of Kowalski et al. (2006), which showed no significant background, despite being at much lower Galactic latitude.  The observations by Murthy et al. show that a similar short-wavelength patchy diffuse background occurs everywhere.    \label{Fig31}}
\end{figure}
\clearpage

\end{document}